\documentclass[twocolumn]{aastex63}
\usepackage{amsmath,amssymb}
\usepackage{graphicx}

\begin{document}

\title{NEO Population, Velocity Bias, and Impact Risk from an ATLAS Analysis}

\author{A. N. Heinze}
\affiliation{Institute for Astronomy, University of Hawaii, 2680 Woodlawn, Honolulu, HI, 96822, USA; aheinze@hawaii.edu}

\author{Larry Denneau}
\affiliation{Institute for Astronomy, University of Hawaii, 2680 Woodlawn, Honolulu, HI, 96822, USA; aheinze@hawaii.edu}

\author{John L. Tonry}
\affiliation{Institute for Astronomy, University of Hawaii, 2680 Woodlawn, Honolulu, HI, 96822, USA; aheinze@hawaii.edu}

\author{Steven J. Smartt}
\affiliation{Astrophysics Research Centre, School of Mathematics and Physics, Queen's University Belfast, Belfast, BT7 1NN, UK}

\author{Nicolas Erasmus}
\affiliation{South African Astronomical Observatory, Cape Town, 7925, South Africa}

\author{Alan Fitzsimmons}
\affiliation{Astrophysics Research Centre, School of Mathematics and Physics, Queen's University Belfast, Belfast, BT7 1NN, UK}

\author{James E. Robinson}
\affiliation{Astrophysics Research Centre, School of Mathematics and Physics, Queen's University Belfast, Belfast, BT7 1NN, UK}

\author{Henry Weiland}
\affiliation{Institute for Astronomy, University of Hawaii, 2680 Woodlawn, Honolulu, HI, 96822, USA; aheinze@hawaii.edu}

\author{Heather Flewelling}
\affiliation{Institute for Astronomy, University of Hawaii, 2680 Woodlawn, Honolulu, HI, 96822, USA; aheinze@hawaii.edu}

\author{Brian Stalder}
\affiliation{Rubin Observatory, 950 N. Cherry Ave, Tucson, AZ 85719, USA}

\author{Armin Rest}
\affiliation{Space Telescope Science Institute, 3700 San Martin Drive, Baltimore, MD 21218, USA}
\affiliation{Department of Physics and Astronomy, Johns Hopkins University, Baltimore, MD 21218, USA}

\author{David R. Young}
\affiliation{Astrophysics Research Centre, School of Mathematics and Physics, Queen's University Belfast, Belfast, BT7 1NN, UK}

\correspondingauthor{A. N. Heinze} 

\begin{abstract}
We estimate the total population of near-Earth objects (NEOs) in the Solar System, using an extensive, `Solar System to pixels' fake-asteroid simulation to debias detections of real NEOs by the ATLAS survey. Down to absolute magnitudes $H=25$ and 27.6 (diameters of $\sim 34$ and 10 meters, respectively, for 15\% albedo), we find total populations of $(3.72 \pm 0.49) \times 10^5$ and $(1.59 \pm 0.45) \times 10^7$ NEOs, respectively. Most plausible sources of error tend toward underestimation, so the true populations are likely larger. We find the distribution of $H$ magnitudes steepens for NEOs fainter than $H \sim 22.5$, making small asteroids more common than extrapolation from brighter $H$ mags would predict. Our simulation indicates a strong bias against detecting small but dangerous asteroids that encounter Earth with high relative velocities --- i.e., asteroids in highly inclined and/or eccentric orbits. Worldwide NEO discovery statistics indicate this bias affects global NEO detection capability, to the point that an observational census of small asteroids in such orbits is probably not currently feasible. Prompt and aggressive followup of NEO candidates, combined with closer collaborations between segments of the global NEO community, can increase detection rates for these dangerous objects.
\end{abstract}


\section{Introduction}

The question of the total population, down to some minimum size, of Solar System objects in various categories is an old one that has been addressed in many different ways, for classes ranging from Oort Cloud objects to the hypothetical (but probably nonexistent) Vulcanoids inside the orbit of Mercury. For some classes, the question can be answered confidently if the minimum size being considered is large enough. The $2.4\times 10^4$ known main belt asteroids larger than 5 km (assuming a 15\% albedo) certainly constitute the vast majority of such objects actually in existence; and similarly the 902 known NEOs larger than 1 km are believed to make up 97\% of the true population above this size limit \citep{Stokes2017}. But in both populations, surveys have detected many thousands of much smaller objects. This presents us with the opportunity to constrain the total population down to a much smaller minimum size -- but such estimates require `debiasing': that is, estimating the fraction of the total population that has been discovered (as a function of asteroid size), and applying the appropriate correction to the counts of known asteroids of various sizes.

\medskip

Accurate debiasing is not an easy task, especially for the NEOs, because current discoveries are the result of many surveys operating with heterogeneous strategies and sensitivities over many years --- and because so many different survey parameters affect discovery rates over time. A further concern is to what extent the properties of the known objects reliably inform us about those that remain to be discovered. In other words, we must engage with the question of whether properties other than size cause nontrivial biases in discovery statistics.

\medskip

The results we present herein are based on data from the Asteroid Terrestrial-impact Last Alert System \citep[ATLAS; ][]{Tonry2018}. ATLAS uses custom-built 0.5 meter F/2 telescopes equipped with 110 megapixel STA-1600 detectors delivering solid angle coverage of 29 square degrees per image at a pixel scale of 1.86 arcsec. Images are acquired in two customized broadband filters: the $o$ band (`orange', $\sim r+i$) for most observations, and the shorter-wavelength $c$ band (`cyan', $\sim g+r$) near new moon on one telescope. A precise photometric calibration is obtained for every image using the ATLAS All-Sky Stellar Reference Catalog \citep{refcat}. In good conditions, ATLAS achieves a magnitude limit of about 19.5 in each 30 second exposure \citep{Smith2020}. Although this magnitude limit is brighter than most surveys, ATLAS covers more solid angle on the sky per night than any other planetary defense survey, and unlike others does not avoid the Galactic plane. This rapid scanning of the whole accessible sky serves ATLAS' mission to function as a `last alert' --- that is, to detect small but dangerous asteroids incoming for impact. Hence, ATLAS is complementary to other planetary defense surveys whose primary mission is to detect larger asteroids decades before they might hit the Earth. Though all NEO surveys can (and do) discover NEOs of all sizes, ATLAS focuses on the smallest cohort of dangerous asteroids: objects like the Tunguska impactor of 1908, which are large enough to destroy cities but small enough that they are usually discovered only when quite close to the Earth.

\medskip

In Section \ref{sec:photan} we present an extensive simulation we performed to debias NEO detections from ATLAS. In Section \ref{sec:totpop}, we use the results to estimate the population of NEOs. Section \ref{sec:bias} describes our finding of a strong bias against detecting small but dangerous (20-100 meter) asteroids that encounter Earth at high relative velocity, and shows that this bias applies not only to ATLAS but to the global NEO discovery capability. In Section \ref{sec:biasrisk}, we explore the implications for impact risk and suggest ways to improve sensitivity to high-velocity asteroids. We offer our conclusions in Section \ref{sec:conc}.

\section{The ATLAS NEO Simulation} \label{sec:photan}

We have carried out an extensive simulation aimed at accurately debiasing ATLAS NEO detections in order to estimate the total number of NEOs. More specifically, we seek to determine the distribution of NEOs as a function of $H$ magnitude (which can be used as a proxy for size; see Table \ref{tab:Hmag}). This distribution can then be integrated to yield an estimate of the total number of NEOs in the Solar System larger than any desired size threshold, down to some minimum size below which our data are no longer adequate.

\begin{deluxetable}{lr}
\tablewidth{0pt}
\tabletypesize{\footnotesize}
\tablecaption{Asteroid Absolute Magnitude and Size\label{tab:Hmag}}
\tablehead{ \colhead{Absolute} & \colhead{Diameter} \\
\colhead{mag. ($H$)} & \colhead{for 15\% albedo}}
\startdata
17.0 & 1400 m  \\
20.0 & 340 m \\
21.0 & 220 m \\
22.0 & 140 m \\
23.0 & 86 m \\
24.0 & 54 m \\
25.0 & 34 m \\
26.0 & 22 m \\
27.0 & 14 m \\
28.0 & 9 m \\
\enddata
\end{deluxetable}

\subsection{Mathematical framework of the simulation}

The observational data we seek to use is $D(H)$, the number of NEOs detected by ATLAS as a function of absolute magnitude $H$, within a specified period of time. Our simulation uses fake asteroids to debias $D(H)$ by determining the function $f_d(H)$ that describes the $H$-dependent fraction of all NEOs that should be detected by ATLAS over the same period. The resulting estimate of $N(H)$, the differential $H$ distribution for all NEOs in the Solar System, is given by:

\begin{equation} \label{eq:diff}
N(H) = \frac{D(H)}{f_d(H)}
\end{equation}

Then the cumulative distribution $N(<H)$ for all NEOs in the Solar System is:

\begin{equation} \label{eq:cum}
N(<H) = \int_{0}^{H} N(H') dH'
\end{equation}

Equation \ref{eq:cum} yields an approximate cumulative size distribution if we convert the threshold absolute magnitude $H$ into a size using an adopted mean albedo for NEOs. Based on the finding of a mean albedo of 0.147 by \citet{Morbidelli2020}, we have uniformly adopted a round number of 15\% herein (e.g. Table \ref{tab:Hmag}), noting that this is also consistent with the taxonomic mix found for NEOs and inner Main Belt asteroids by \citet{Erasmus2017}; \citet{Mommert2016}; and \citet{Erasmus2018}.

\subsection{Simulation Design}

The purpose of our simulation is to determine $f_d(H)$, the fraction of all NEOs with absolute magnitude $H$ that are expected to be detected by ATLAS over a given time period. To achieve this, we use an ambitious `Solar System to pixels' approach. The chosen time period is an approximately 15-month span, from 2017 June 01 through 2018 August 22\footnote{The beginning and end of this time period correspond approximately to the completion of improvements to both ATLAS telescopes (including the replacement of the Schmidt correctors and fine-tuning of the collimation), and to the real-world date on which we launched the simulation, respectively.}.
The outline of the simulation is as follows:

\begin{itemize}

\item Create a list of orbits for an `overpopulated' Solar System containing hundreds of times more NEOs than are actually thought to exist.

\item Turn the simulated orbits into simulated asteroids by assigning a specific $H$ magnitude to each one, defining in the process a simulated input $H$ magnitude distribution $S(H)$.

\item Cull out the vast majority of simulated NEOs that remain much fainter than the ATLAS detection limit throughout the period being simulated.

\item Calculate precise orbital ephemerides for the remaining simulated NEOs, which had some chance of being detectable by ATLAS.

\item Match these ephemerides with ATLAS images, and, where appropriate, `paint' the fake NEOs into the images with the correct positions, magnitudes\footnote{Rotational flux variations (lightcurves) are ignored, and all asteroids are assumed to have the same color and phase function; see further discussion in Sections \ref{sec:simdetail} and \ref{sec:approx}.}, and trail lengths.

\item Process these `faker' images using the same analysis ATLAS uses to detect real asteroids, including the final step of `linking' individual detections into `tracklets' each comprising a set of measurements of a particular object on a particular night.

\item Construct the list of simulated NEOs that produced at least one tracklet during the period of the simulation. Only these objects count as being `recovered'.

\item Construct the $H$ magnitude distribution $R(H)$ for the simulated NEOs that were recovered, and divide it by the input distribution $S(H)$ to obtain the detection fraction $f_d(H)$.

\end{itemize}

Hence, the simulation's final product is given by:

\begin{equation} \label{eq:detfrac}
f_d(H) = \frac{R(H)}{S(H)}
\end{equation}

Where $S(H)$ is the total number of NEOs of absolute magnitude $H$ in the simulation's `overpopulated' Solar System; and $R(H)$ is the number of distinct simulated NEOs that were recovered by the ATLAS detection pipeline. Recovery here requires being detected enough times on a single night that the Moving Object Processing System \citep[MOPS; ][]{Denneau2013} can confidently link the successive detections into a tracklet. This is also the criterion for submitting a real asteroid to the Minor Planet Center (MPC). In general, ATLAS requires tracklets to have at least four detections --- a constraint that is necessary in order to avoid excessive false positives. In the case of fast-moving objects whose images are trailed, ATLAS allows tracklets with only three detections: for such objects, MOPS can eliminate most false positives by requiring all three detections to be trails with consistent orientation. Our simulation captures both of these discovery pathways. 

\medskip

Given the overpopulated Solar System and corresponding vast number of tracklets produced by MOPS, manual screening of simulated asteroid discoveries was not practical --- although real ATLAS discoveries are manually screened before submission to the MPC. The realism of the faker simulation depends on the assumption that this manual screening would not have resulted in the rejection of any significant fraction of the tracklets corresponding to `real' fake asteroids. On a typical night of real ATLAS data, roughly 90\% of tracklets are rejected as spurious by whichever ATLAS team member is tasked with the manual screening for that night. In nearly all cases the bad tracklets are composed of obviously spurious detections such as cosmic rays, diffraction spikes, and ghost reflections from bright stars, so the decision to reject them is an easy one. Given this, the assumption that very few of the tracklets corresponding to simulated asteroids would have been manually rejected seems to be fairly safe. Note that the violation of this assumption would cause overestimation of the detection fraction $f_d(H)$, and would therefore lead to undercounting of the true number of asteroids in the Solar System. As we discuss in more detail below, this tendency to undercounting is shared by almost all other types of error that can affect our analysis: hence, the final population estimates are something in the nature of lower limits.

\subsection{Details of the Simulation} \label{sec:simdetail}

The input $H$ magnitude distribution $S(H)$ used for our simulation is described in Table \ref{tab:inputcounts}. It is intended not to approximate the $H$ magnitude distribution of real NEOs, but rather to over-represent small NEOs in order to ensure that the simulation has sufficient statistical power to constrain $f_d(H)$ even when it is very small -- i.e., for tiny, hard-to-detect asteroids. The fact that $S(H)$ does not match the absolute magnitude distribution of real NEOs does not affect the accuracy of $f_d(H)$, since the differences cancel in the ratio on the right hand side of Equation \ref{eq:detfrac}.

\medskip

\begin{deluxetable}{llc}
\tablewidth{0pt}
\tabletypesize{\footnotesize}
\tablecaption{Absolute Magnitude Distribution of Simulated Asteroids\label{tab:inputcounts}}
\tablehead{ &  & \colhead{Number} \\
\colhead{H mag range} & \colhead{H mag distribution} & \colhead{Simulated}}
\startdata
$17.0 < H < 25.0$ & Follows \citet{Granvik2018} & $2\times10^8$ \\
$25.0 < H < 26.0$ & Uniform & $6\times10^9$ \\
$26.0 < H < 27.0$ & Uniform & $6\times10^9$ \\
$27.0 < H < 28.0$ & Uniform & $10^{10}$ \\
$28.0 < H < 29.0$ & Uniform & $10^{10}$ \\
$29.0 < H < 30.0$ & Uniform & $10^{10}$ \\
Total & \nodata & $4.22 \times 10^{10}$ \\
\enddata
\end{deluxetable}

Unique orbits for the $4.22 \times 10^{10}$ simulated NEOs described in Table \ref{tab:inputcounts} were obtained from the Granvik model \citep{Granvik2018} by cloning the Keplerian orbital elements of the $8 \times 10^5$ NEOs included in the model. To clone an orbit, we adopted the three most physically significant elements (semimajor axis $a$, eccentricity $e$, and inclination $i$) without alteration, but selected the parameters defining the orientation of the orbit (mean anomaly, argument of perihelion, and longitude of the ascending node) randomly from a uniform distribution on 0--360$^{\circ}$. We used this approach because the orbital distributions of real asteroids can exhibit complex and highly significant correlations between $a$, $e$, and $i$, but the distributions of the other orbital elements are nearly uniform and any correlations are weak. The $H$ magnitudes in the range covered by the Granvik model ($H \leq 25.0$) were randomly drawn from the magnitudes in the model. This random selection of $H$ magnitudes was independent of the random drawing of orbits, so that a given combination of $a$, $e$, and $i$ would be paired with many different magnitudes. For asteroids fainter than $H=25.0$, simulated $H$ magnitudes were chosen randomly from uniform distributions as indicated by Table \ref{tab:inputcounts}.

\medskip

This process for assigning simulated $H$ magnitudes carries an implicit assumption that the distribution of NEO orbital elements does not depend on $H$ magnitude --- in other words, that the orbits of small NEOs are distributed identically to the orbits of large ones. This assumption is certainly not true in detail (the \citet{Granvik2018} model itself contains slight departures from it, which are obliterated by our process of randomly re-assigning $H$ magnitudes), but we believe it is a reasonable approximation given the present state of knowledge. The process that moves asteroids inward from the Main Belt to become NEOs is believed to involve the Yarkovsky effect, which is size-dependent \citep{Farinella1998,Nesvorny2004} --- but the largest orbital changes are produced by size-independent causes, including resonant interactions with the Jovian planets and gravitational scattering by the terrestrial worlds \citep[e.g.][]{Bottke2002}. Hence, it is not unreasonable to suppose that the orbital distributions of large and small NEOs would be approximately the same. In addition (as we demonstrate below), observational data are currently insufficient to accurately determine the orbital distribution of small NEOs, so we {\em might as well} assume it is the same as for the larger objects.

\medskip

The orbits of all $4.22 \times 10^{10}$ simulated NEOs were coarsely integrated with a 24 hr time step to identify those that became brighter than apparent magnitude 20.0 {\em or} approached within 0.02 AU of the Earth regardless of their apparent magnitude. The 20.0 magnitude limit is intended capture all asteroids that could conceivably have been detected given the ATLAS telescopes' point-source sensitivity limit of about magnitude 19.5. The 0.02 AU threshold is meant to catch smaller asteroids with close encounters, whose brief period of detectability could fall between two 24 hr time steps. Only $6.58 \times 10^6$ simulated NEOs, or 0.016\% of our input objects, survived this preliminary culling for plausible detectability. 

\medskip

The orbit of each plausibly detectable NEO was precisely integrated, and its position evaluated at the time of each ATLAS image during the 15 months spanned by the simulation. Simulated asteroids whose ephemerides placed them on an ATLAS image were `painted' into the image with sub-arcsecond precision. The point spread function (PSF) used for inserting these fake asteroids was based on a set of template stars on the same image; and simulated NEOs with fast angular velocities were given correspondingly trailed images at the correct orientation.

\medskip

\begin{figure*}
\includegraphics[width=7.0in]{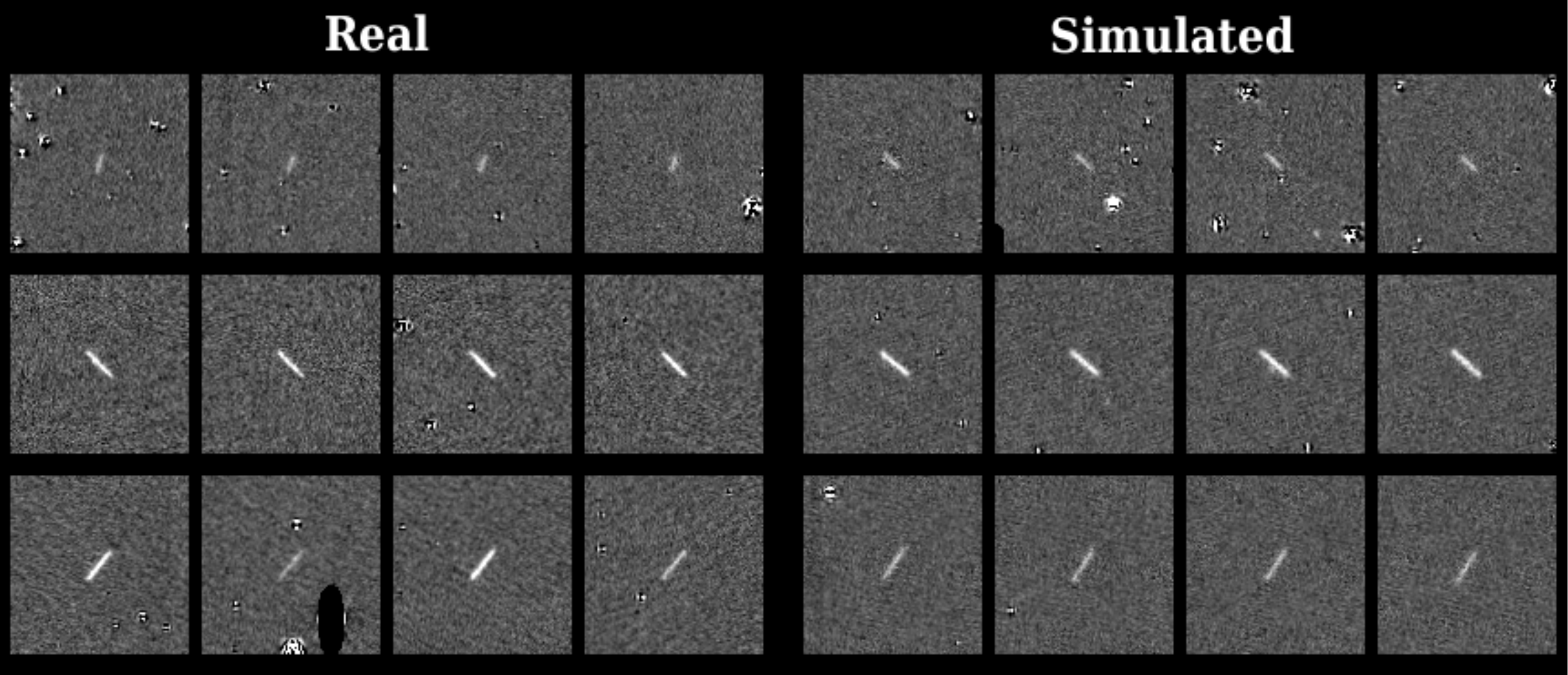}
\caption{Examples of three real (left) and three simulated (right) fast-moving asteroids imaged by ATLAS. Each object is detected four times over a period of about forty minutes, following ATLAS' typical observing pattern. The three real asteroids are all ATLAS discoveries: from top to bottom they are 2020 BF6, 2019 TN5, and 2020 OH. The fake asteroids are nearly indistinguishable from real ones, though we do not simulate rotationally modulated brightness variation such as is apparently shown by 2020 OH in these images.
\label{fig:fakers}}
\end{figure*}

The painting of simulated asteroids into actual image pixels concludes the setup of our simulation. Its `Solar System to pixels' approach implicitly includes the effects of weather, moonlight, and equipment problems as they applied to real survey images used to detect real asteroids. There is no need for a statistical model of weather or other effects, because our simulated asteroids encounter the same actual weather as the real NEOs detected by ATLAS over the same period. They experience, image by image, the same changes in sky transparency, sky background, and PSF sharpness that affect the detections of real objects. Figure \ref{fig:fakers} compares real and `painted' asteroids, illustrating the pixel-level realism of our fake objects.

\medskip

Our survey accounts for another distinct set of effects that can prevent the discovery of real NEOs by surveys: effects relating to their angular motion on the sky. High angular velocities can make the trailed images of NEOs too long and faint and/or their successive detections too far apart for the linking software, causing them to go undetected even at relatively bright magnitudes. They can also be missed due to excessive angular {\em acceleration}: a curved and/or linearly accelerating trajectory across the sky may depart too much from a constant-velocity Great Circle for the linking program (e.g. MOPS) to recognize multiple consecutive detections of the same object. The rotation of the Earth is the dominant cause of curvature in an asteroid's on-sky track, while a rapid fractional change in distance is the main cause of acceleration along a Great Circle. All of these effects are strongest at small distances from the Earth. Hence, they have the greatest effect on the smallest asteroids, which can only be detected when close. By integrating physically realistic orbits and calculating precise topocentric ephemerides, we ensure that the fake asteroids in the ATLAS NEO simulation are subject to the same linkage losses as real objects would be.

\medskip

One characteristic of real NEOs that was {\em not} explicitly included in the ATLAS simulation is rotational brightness variation. This can be sufficient to prevent an object's discovery. For example, if the mean magnitude of an asteroid is 19.2 and its brightness varies with a range of one magnitude, it is quite likely to be fainter than ATLAS' 19.5 magnitude detection limit on one or more of the four images required for a discovery. Some known NEOs have variations quite a bit larger than this. Hence, the simulation may overestimate the detection fraction if rotational brightness variations are a significant cause of lost discoveries.

\medskip

\begin{figure}
\includegraphics[width=3.5in]{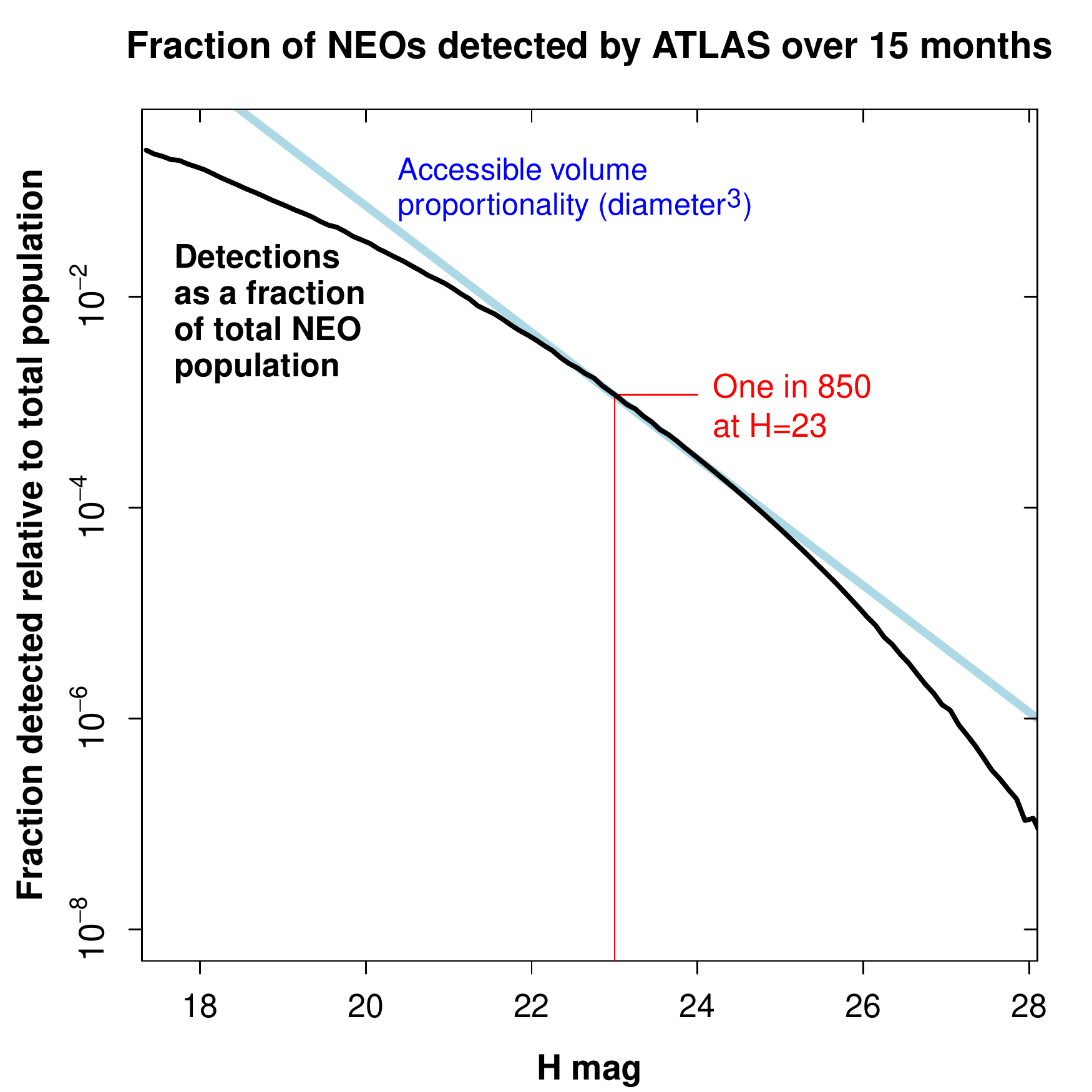}
\caption{The fractional detection curve $f_d(H)$ from the ATLAS NEO simulation. The example value of $f_d(23) = \frac{1}{850}$ indicates that for every real NEO of $H=23$ detected by ATLAS during the period covered by the simulation, we can infer that 850 such objects actually exist in the Solar System. The pale blue curve illustrates the idealized $f_d(H)$ that would result if NEOs were evenly distributed throughout an evenly illuminated three-dimensional space. The curve shown here is $f_d(H_V)$: it has been corrected for the mean difference between the ATLAS $o$ and $c$ bands and the standard $V$ band $H$ mags used by the MPC.
\label{fig:fd}}
\end{figure}

The fractional detection curve $f_d(H)$ from the ATLAS NEO simulation is shown in Figure \ref{fig:fd}. It is interesting to compare it with a power law proportional to the cube of the asteroid diameter, which would match the fractional detection curve if NEOs were evenly illuminated and evenly distributed through three-dimensional space. The actual curve falls below this idealized situation for large asteroids because NEOs are concentrated near the ecliptic and are less illuminated when far from the Sun; and for small asteroids because of the angular velocity losses and other effects already discussed.

\medskip

In summary, the salient characteristics of the ATLAS NEO simulation are as follows:

\begin{itemize}

\item Aimed at obtaining the fractional detection curve $f_d(H)$ as a function of absolute magnitude $H$.

\item Self-consistent modeling from the whole inner Solar System down to image pixels.

\item Statistically powerful, simulating hundreds of times more NEOs than are actually believed to exist.

\item Used the approximation that small NEOs have the same orbital distribution as large ones.

\item Realistic enough to include most (though not all) of the effects that can prevent the discovery of real asteroids.

\end{itemize}

\section{Total NEO Population from the ATLAS Simulation} \label{sec:totpop}

Our simulation included artificial NEOs with $H$ magnitudes ranging from the bright limit of the Granvik model ($H=17.0$) down to $H=30.0$, producing the fractional detection curve $f_d(H)$ shown in Figure \ref{fig:fd}. During the 15-month period covered by the simulation, ATLAS detected 713 distinct real NEOs with absolute magnitudes in the regime probed by the simulation (the faintest ATLAS detection was 2017 SU$_{17}$ at $H=28.0$). Of these 713 distinct NEOs, 140 (including 2017 SU$_{17}$) were ATLAS discoveries, while the rest were independent recoveries of NEOs discovered elsewhere. The absolute magnitude distribution of real detections (called $D(H)$ in Equation \ref{eq:diff}) is simply the histogram of these 713 detected NEOs. It is plotted in the left panel of Figure \ref{fig:diffhist}.

\subsection{Approximations involved in debiasing $D(H)$} \label{sec:approx}

The histogram $D(H)$ is based on absolute magnitudes calculated and cataloged by the MPC, which are standardized to the $V$ band: hence, we can refer to it more specifically as $D(H_V)$, while the $f_d(H)$ curve from our simulation is based on the ATLAS $o$ and $c$ bands rather than the $V$ band. A color correction\footnote{Ideally, this correction would have been built into our simulation from the beginning, but we were not aware of its importance until late in our analysis and therefore performed the correction in an average sense after the fact.} must therefore be applied to our $f_d(H)$ curve to make it match $D(H_V)$. The MPC uses mean asteroid colors, based on ATLAS measurements, to correct $o$ and $c$ band measurements to standard $H_V$ mags: $<V-o> = +0.332$ and $<V-c> = -0.054$. Most ATLAS images are in the $o$ band, and in fact only 14.6\% of our simulated asteroids were detected in $c$ band images. Hence, we derive a correction of 0.276 mag: the mean of the $V-c$ and $V-o$ colors weighted by their respective fractional contributions of 0.146 and 0.854.  The correction is in the sense that the average ATLAS measurement is brighter than the corresponding $V$ magnitude: $H_{ATLAS} = H_V - 0.276$. This correction has the effect of reducing our calculated NEO populations below the values that would result from the faulty assumption that $H_{ATLAS}=H_V$. For example, as we debias using Equation \ref{eq:diff}, $D(H_V = 27.0)$ gets scaled by $f_d(H_{ATLAS}=26.724)^{-1}$, which is smaller than the scaling factor $f_d(H_{ATLAS}=27.0)^{-1}$ that would have been used apart from the correction. From this point forward, we will use $H$ without a subscript to refer to the MPC standard $H_V$.

\medskip

Our analysis, and indeed the MPC $H$ magnitudes on which it relies, implicitly assume that all NEOs are the same color. This is just one of several approximations that are common to the field but not always explicitly mentioned. Another is the MPC's use of the \citet{Bowell1989} phase function with fixed slope parameter $G=0.15$ for most NEOs in deriving $H$ magnitudes from reported apparent magnitudes. Since the $H$ magnitude is defined at phase=0, while most observations are obtained at nonzero phase, determining $H$ from observed magnitudes depends on the phase function. Where sufficient photometry exists to solve for object-specific phase slopes (mostly for Main Belt asteroids), considerable variation is seen in the best-fit values of $G$ for different asteroids \citep{Pravec2012,Veres2015}: hence the assumption of $G=0.15$ can introduce errors into the $H$ magnitudes of NEOs. A further approximation --- that of a constant mean albedo --- comes into play whenever we convert $H$ magnitudes into size. Like our own approximation that the orbital distribution of NEOs is independent of size, these other widely-adopted approximations are enforced by the lack of data sufficient to support more realistic models. Deriving individualized colors, phase functions, and albedos for every NEO that is discovered would require intensive, multi-wavelength observations over a long temporal arc, as well as the accurate modeling of rotational variations. These data do not exist for most small NEOs --- although current and future programs of intensive observations and analysis may ultimately enable more sophisticated statistical constraints.

\medskip

Subject to all these approximations, dividing $D(H)$ by the color-corrected fractional detection curve $f_d(H)$ produces our estimate of $N(H)$, the absolute magnitude histogram for all NEOs in the Solar System. The left panel of Figure \ref{fig:diffhist} illustrates Equation \ref{eq:diff} in practice, including $D(H)$; the scaling factor $f_d(H)^{-1}$; and the corrected histogram $N(H)$. Following Equation \ref{eq:cum}, we integrate $N(H)$ to produce the cumulative distribution, adding in the probably complete count of 600 known NEOs brighter than $H=17.3$. The result is plotted in the right panel of Figure \ref{fig:diffhist}. Table \ref{tab:NEOpop} presents $f_d(H)$, $N(H)$, and the cumulative distribution in numerical form, using coarser sampling than the figure to avoid an excessively long table.

\begin{deluxetable*}{ccccccccc}
\tablewidth{0pt}
\tabletypesize{\footnotesize}
\tablecaption{ATLAS Results on the NEO Population\label{tab:NEOpop}}
\tablehead{\colhead{H mag at} & \colhead{detected} & & \colhead{uncertainty}  & \colhead{total NEOs:} & \colhead{uncertainty} & \colhead{cumulative} & \colhead{uncertainty} & \colhead{cum. upper} \\
\colhead{bin center} & \colhead{NEOs: $D(H)$} & \colhead{$f_d(H)$} & \colhead{on $f_d(H)$} & \colhead{$N(H)$} & \colhead{on $N(H)$} & \colhead{NEOs: $N(<H)$} & \colhead{on $N(<H)$} & \colhead{envelope}}
\startdata
17.4 & 21 & 2.315E-001 & 4.239E-003 & 9.070E+001 & 1.985E+001 & 6.327E+002 & 1.158E+001 & \nodata \\
17.6 & 18 & 2.034E-001 & 3.727E-003 & 8.847E+001 & 2.090E+001 & 7.189E+002 & 2.294E+001 & \nodata \\
17.8 & 14 & 1.867E-001 & 3.072E-003 & 7.497E+001 & 2.007E+001 & 8.099E+002 & 3.144E+001 & \nodata \\
18.0 & 20 & 1.639E-001 & 2.615E-003 & 1.220E+002 & 2.733E+001 & 9.011E+002 & 3.886E+001 & \nodata \\
18.2 & 24 & 1.407E-001 & 1.888E-003 & 1.706E+002 & 3.489E+001 & 1.065E+003 & 5.085E+001 & \nodata \\
18.4 & 37 & 1.193E-001 & 1.663E-003 & 3.101E+002 & 5.110E+001 & 1.300E+003 & 6.661E+001 & \nodata \\
18.6 & 28 & 1.009E-001 & 1.455E-003 & 2.776E+002 & 5.257E+001 & 1.561E+003 & 8.246E+001 & \nodata \\
18.8 & 29 & 8.664E-002 & 1.266E-003 & 3.347E+002 & 6.231E+001 & 1.889E+003 & 1.014E+002 & \nodata \\
19.0 & 25 & 7.440E-002 & 1.053E-003 & 3.360E+002 & 6.734E+001 & 2.237E+003 & 1.210E+002 & \nodata \\
19.2 & 31 & 6.324E-002 & 8.942E-004 & 4.902E+002 & 8.823E+001 & 2.587E+003 & 1.405E+002 & \nodata \\
19.4 & 32 & 5.451E-002 & 8.049E-004 & 5.871E+002 & 1.041E+002 & 3.211E+003 & 1.742E+002 & \nodata \\
19.6 & 21 & 4.640E-002 & 6.924E-004 & 4.526E+002 & 9.890E+001 & 3.693E+003 & 2.001E+002 & \nodata \\
19.8 & 21 & 3.837E-002 & 6.174E-004 & 5.474E+002 & 1.198E+002 & 4.104E+003 & 2.225E+002 & \nodata \\
20.0 & 24 & 3.339E-002 & 5.577E-004 & 7.187E+002 & 1.471E+002 & 4.918E+003 & 2.692E+002 & \nodata \\
20.2 & 27 & 2.678E-002 & 5.168E-004 & 1.008E+003 & 1.948E+002 & 5.518E+003 & 3.042E+002 & \nodata \\
20.4 & 19 & 2.231E-002 & 4.627E-004 & 8.517E+002 & 1.960E+002 & 6.463E+003 & 3.606E+002 & \nodata \\
20.6 & 18 & 1.864E-002 & 4.612E-004 & 9.658E+002 & 2.285E+002 & 7.520E+003 & 4.257E+002 & \nodata \\
20.8 & 18 & 1.517E-002 & 3.870E-004 & 1.186E+003 & 2.807E+002 & 8.530E+003 & 4.917E+002 & \nodata \\
21.0 & 16 & 1.254E-002 & 3.622E-004 & 1.275E+003 & 3.204E+002 & 9.737E+003 & 5.728E+002 & \nodata \\
21.2 & 16 & 1.015E-002 & 3.515E-004 & 1.577E+003 & 3.967E+002 & 1.143E+004 & 6.941E+002 & \nodata \\
21.4 & 17 & 7.912E-003 & 3.146E-004 & 2.149E+003 & 5.253E+002 & 1.341E+004 & 8.463E+002 & \nodata \\
21.6 & 11 & 6.497E-003 & 3.128E-004 & 1.693E+003 & 5.146E+002 & 1.509E+004 & 9.766E+002 & \nodata \\
21.8 & 12 & 5.110E-003 & 2.849E-004 & 2.349E+003 & 6.855E+002 & 1.721E+004 & 1.156E+003 & \nodata \\
22.0 & 8 & 4.150E-003 & 2.605E-004 & 1.928E+003 & 6.881E+002 & 1.941E+004 & 1.353E+003 & \nodata \\
22.2 & 12 & 3.369E-003 & 2.716E-004 & 3.562E+003 & 1.055E+003 & 2.310E+004 & 1.712E+003 & \nodata \\
22.4 & 8 & 2.466E-003 & 2.389E-004 & 3.245E+003 & 1.171E+003 & 2.527E+004 & 1.935E+003 & \nodata \\
22.6 & 8 & 1.971E-003 & 2.232E-004 & 4.059E+003 & 1.475E+003 & 2.889E+004 & 2.337E+003 & \nodata \\
22.8 & 7 & 1.608E-003 & 2.075E-004 & 4.354E+003 & 1.705E+003 & 3.401E+004 & 2.930E+003 & \nodata \\
23.0 & 11 & 1.209E-003 & 1.772E-004 & 9.099E+003 & 2.931E+003 & 4.177E+004 & 3.940E+003 & \nodata \\
23.2 & 14 & 9.075E-004 & 1.589E-004 & 1.543E+004 & 4.553E+003 & 5.295E+004 & 5.408E+003 & \nodata \\
23.4 & 13 & 6.827E-004 & 1.359E-004 & 1.904E+004 & 5.950E+003 & 6.813E+004 & 7.250E+003 & \nodata \\
23.6 & 7 & 5.212E-004 & 1.182E-004 & 1.343E+004 & 5.524E+003 & 8.630E+004 & 9.502E+003 & \nodata \\
23.8 & 10 & 3.953E-004 & 9.344E-005 & 2.530E+004 & 9.036E+003 & 1.041E+005 & 1.186E+004 & \nodata \\
24.0 & 5 & 3.083E-004 & 6.204E-005 & 1.622E+004 & 7.778E+003 & 1.303E+005 & 1.527E+004 & \nodata \\
24.2 & 15 & 2.163E-004 & 4.235E-005 & 6.934E+004 & 2.035E+004 & 1.592E+005 & 1.945E+004 & \nodata \\
24.4 & 5 & 1.647E-004 & 2.763E-005 & 3.035E+004 & 1.410E+004 & 2.203E+005 & 2.770E+004 & \nodata \\
24.6 & 8 & 1.237E-004 & 1.715E-005 & 6.466E+004 & 2.408E+004 & 2.801E+005 & 3.563E+004 & \nodata \\
24.8 & 4 & 8.997E-005 & 1.006E-005 & 4.446E+004 & 2.267E+004 & 3.302E+005 & 4.237E+004 & \nodata \\
25.0 & 6 & 6.042E-005 & 5.521E-006 & 9.930E+004 & 4.114E+004 & 3.722E+005 & 4.896E+004 & 5.64E+005 \\
25.2 & 6 & 4.552E-005 & 2.825E-006 & 1.318E+005 & 5.437E+004 & 5.247E+005 & 7.340E+004 & 7.93E+005 \\
25.4 & 6 & 3.205E-005 & 1.238E-006 & 1.872E+005 & 7.695E+004 & 6.909E+005 & 1.003E+005 & 1.07E+006 \\
25.6 & 2 & 1.993E-005 & 7.556E-007 & 1.004E+005 & 7.104E+004 & 7.610E+005 & 1.119E+005 & 1.31E+006 \\
25.8 & 5 & 1.460E-005 & 4.822E-007 & 3.425E+005 & 1.540E+005 & 9.827E+005 & 1.580E+005 & 1.77E+006 \\
26.0 & 2 & 9.116E-006 & 3.957E-007 & 2.194E+005 & 1.554E+005 & 1.204E+006 & 2.032E+005 & 2.37E+006 \\
26.2 & 7 & 6.316E-006 & 3.978E-007 & 1.108E+006 & 4.260E+005 & 1.686E+006 & 3.163E+005 & 3.39E+006 \\
26.4 & 2 & 4.026E-006 & 4.080E-007 & 4.968E+005 & 3.545E+005 & 2.532E+006 & 4.969E+005 & 5.12E+006 \\
26.6 & 4 & 2.948E-006 & 3.422E-007 & 1.357E+006 & 6.920E+005 & 3.629E+006 & 7.464E+005 & 7.53E+006 \\
26.8 & 3 & 1.963E-006 & 3.039E-007 & 1.528E+006 & 9.021E+005 & 5.335E+006 & 1.149E+006 & 1.15E+007 \\
27.0 & 1 & 1.195E-006 & 2.999E-007 & 8.365E+005 & 8.603E+005 & 5.913E+006 & 1.290E+006 & 1.55E+007 \\
27.2 & 2 & 7.751E-007 & 1.441E-007 & 2.580E+006 & 1.863E+006 & 7.897E+006 & 1.946E+006 & 2.28E+007 \\
27.4 & 3 & 4.597E-007 & 3.575E-008 & 6.526E+006 & 3.811E+006 & 1.114E+007 & 3.030E+006 & 3.47E+007 \\
27.6 & 0 & 2.946E-007 & 2.287E-008 & 0.000E+000 & 0.000E+000 & 1.586E+007 & 4.517E+006 & 5.60E+007 \\
27.8 & 2 & 1.899E-007 & 1.833E-008 & 1.053E+007 & 7.525E+006 & 2.057E+007 & 6.542E+006 & 8.19E+007 \\
28.0 & 1 & 1.130E-007 & 1.397E-008 & 8.850E+006 & 8.920E+006 & 2.639E+007 & 8.776E+006 & 1.47E+008 \\
\enddata
\end{deluxetable*}

\subsection{Survey Strategy and Simulation Realism}

The scaling of real NEO counts by simulated $f_d(H)$ (Equation \ref{eq:diff}) produces accurate results only to the extent that the detections of real asteroids are obtained using the same criteria as the simulated asteroids. In our simulation, asteroids were detected blindly based on automatically-linked tracklets produced by MOPS. They were only identified with input asteroids after the detection process was complete. Hence, any detections of real asteroids that were {\em not} blind, automatic, and tracklet-based must not be counted in calculating $D(H)$. One example of such a non-blind detection would be targeted followup of an asteroid recently discovered by another survey. A real asteroid detected this way would have an unfair advantage over its simulated brethren because we altered our observations specifically to find it. Another example would be precoveries or recoveries that could be extracted from ATLAS images only with the aid of ephemerides based on non-ATLAS data. Conveniently for our results, ATLAS does not as a rule carry out either targeted followup or ephemeris-based (p)recovery of non-ATLAS discoveries. It is designed to rapidly and impartially survey the whole accessible sky, adhering to an optimized survey pattern except under extraordinary circumstances. Hence, very few detections of real NEOs had to be excluded from our construction of $D(H)$.

\subsection{The Effect of Lost NEOs} \label{sec:lost}

During the period spanned by the simulation, ATLAS detected about twenty NEO candidates that were never successfully followed up by other observatories, nor recovered by ATLAS. Without observations spanning multiple nights, the orbits --- and hence, the absolute magnitudes --- of these `lost' objects cannot be calculated. Since their $H$ magnitudes are not known, they cannot be included in $D(H)$, our histogram of detected real NEOs. 

\medskip

It is possible that some of these `lost' tracklets were spurious, while others might have corresponded to artificial satellites (though none were identified as such by the MPC). We have carefully re-screened all of the lost NEOs to probe these possibilities. This screening included re-examination of the images by ATLAS personnel with experience of tens of thousands of ATLAS tracklets; and also the application of PUMA (Position Using Motion and Acceleration), an orbit-fitting program written by J.T. for rapid dynamical evaluation of ATLAS tracklets. PUMA frequently ruled out a geocentric orbital solution. In cases of objects moving fast enough to be trailed, visual examination always revealed a sequence of four or more trailed detections with trail length and orientation consistent with the on-sky trajectory --- a scenario that is vanishingly unlikely to happen for spurious detections. Rarely was there any evidence of variable stars, cosmic rays, or other sources of spurious detections. Aggressive culling of suspect tracklets only reduced the count of lost objects from $\sim 20$ to 13.

\medskip

These 13 lost NEOs mean that our detection histogram $D(H)$ underestimates the true number of detected NEOs. In the simulation, detections analogous to the lost objects {\em would} be folded into calculating the detection fraction $f_d(H)$, so the fractional detection curve does {\em not} take the loss of these objects into account. Hence, our calculation of $N(H)$, the total population of NEOs, will be underestimated due to the lost objects.

\medskip

At first glance, this does not seem likely to produce a significant error: only 13 objects lost as compared to 713 distinct NEO detections. However, the fast angular velocities and accelerating on-sky trajectories of many of the lost objects suggest they were very close to Earth: hence, they were probably among the smallest NEOs detected by ATLAS during this period. This means they would fall into the least-populated, largest-$H$ bins of our detection histogram, which get maximally amplified in the division by $f_d(H)$. In fact, there are only 14 NEOs in the faintest 1.5 magnitudes of our $D(H)$ histogram. Hence, the 13 lost NEOs might raise our estimate of the total population of small asteroids considerably if they had been recovered and their orbits and $H$ magnitudes had been calculated. Since this didn't happen, all we can do is set an upper limit on how much they could have changed our estimate of the total NEO population.

\medskip

We calculate this upper limit in the context of the cumulative distribution $N(<H)$. To avoid overstating the strength of our constraints, we calculate a very permissive upper limit using the implausible scenario that all 13 of the lost NEOs fall into the faintest absolute magnitude bin, and hence get the largest possible scaling when multiplied by $f_d(H)^{-1}$. For example, the upper limit for $N(<27.0)$ is calculated assuming all 13 objects fell into the $H=27$ bin, and for $N(<28.0)$ we assume they all fell into the $H=28$ bin. The resulting `upper envelope' is plotted as a heavy gray curve in Figure \ref{fig:cumhist}. We emphasize that this is a {\em very} generous upper limit, and that the true NEO population almost certainly falls below it.

\subsection{How to Make a Survey Easy to Simulate} \label{sec:simeasy}

The foregoing discussion of survey realism and lost NEOs offers indications of how an asteroid survey can be made amenable to quantitative simulations such as we have applied to the ATLAS data. First, data processing and NEO detection should be automated and homogeneous, so that the same processing used to detect real asteroids in a given night's data can be applied later to detect simulated NEOs painted into the same data. Exceptions to these requirements, such as followup observations targeting NEOs discovered elsewhere; or ephemeris-dependent (p)recoveries of such asteroids, should be set aside to be excluded from the detection histogram $D(H)$ that will ultimately be processed by the fractional detection curve $f_d(H)$ obtained from the simulation.

\medskip

On the other hand, every effort should be made to avoid the loss of new NEOs detected by the survey. Each object lost without a calculated absolute magnitude is a wild-card that introduces more uncertainty into the final population. The statistical damage is severe enough (e.g. the gray curve in Figure \ref{fig:cumhist}) that heroic efforts may be warranted to prevent the loss of new independently detected NEOs. However, the single most important step is simple: prompt submission of newly detected asteroids to the MPC --- especially if they are fast-moving. The sooner such an object is posted to the MPC's near-Earth Object Confirmation Page (NEOCP), the smaller the ephemeris uncertainty and the more likely it is to be recovered by other observatories \citep{Veres2018}. 

\medskip

Beyond prompt submission, survey self-followup can additionally reduce the risk of lost objects. ATLAS routinely performs same-night self-followup, in which additional observations are scheduled targeting a fast moving, previously unknown NEO. When successful, these increase the temporal arc typically from 30 minutes to 2 hours. In the event that no other observatory recovers the NEO on its discovery night, the lengthened arc greatly reduces the ephemeris uncertainty and increases the chance of recovery on subsequent nights. Unfortunately, same-night self-followup is often impossible: the object might have been discovered too close to sunrise or too low in the west. Under some circumstances, ATLAS will attempt second-night self-followup. Typically in these cases we judge that the object has ephemeris uncertainty so large that most observers in the global followup community will be unwilling to spend valuable observing time on it --- but still small enough that ATLAS, with its huge field of view, has a chance of recovery. Unlike targeted followup of objects discovered elsewhere, these aggressive interventions to recover ATLAS NEOs tend to {\em increase} the fidelity of a potential future simulation rather than detracting from it. The two new ATLAS units currently being constructed in South Africa and Chile will reduce the risk of lost NEOs and the need for second-night self-followup, since most regions of the sky within $40^{\circ}$ of the celestial equator will be covered every 24 hours by the regular survey pattern, as compared to every 48 hours at present.

\medskip

In the long run, loss of NEOs can be reduced by faster submission of all NEO detections, closer cooperation between the respective NEO surveys (as well as between the surveys and the global followup community), and the building of more telescopes in geographically diverse locations. Ironically, greater cooperation between surveys likely means (among other things) inter-survey targeted followup --- one of the things that makes a survey harder to simulate! The solution is simply to keep a strict accounting of NEO detections that were achieved only through external information, and exclude them from the detection histogram $D(H)$ used in Equation \ref{eq:diff}.

\subsection{Discussion of Total NEO Population Results} \label{sec:popdisc}

Figure \ref{fig:cumhist} compares our results on the cumulative distribution of NEO absolute magnitudes with those of several other recent publications. All of them used some type of simulation to model survey performance, but they adopted a variety of different approaches and input data sets. \citet{Harris2015}, \citet{Stokes2017}, and \citet{Tricarico2017} combine results from many surveys over an extended period of time, while \citet{Lilly2017} and \citet{Trilling2017} focus on specific results from Pan-STARRS and DECam, respectively. 

\medskip

Our results are independent of previous work in terms of the input data, and unique in terms of the simulation strategy: we are the first to use ATLAS data and also the first to adopt a comprehensive `Solar System to pixels' approach to calculating the fractional detection curve $f_d(H)$. As illustrated by Figure \ref{fig:cumhist}, we are broadly in agreement with previous observational results both in the approximate total population of NEOs, and in the existence of a change in power law slope between $H=22$ and 23. We note that the model of \citet{Granvik2018} also shows this slope change (Figure \ref{fig:diffhist}). This agreement is significant: although we took the {\em orbital} distribution of our simulated asteroids from the Granvik model, nothing about our simulation would push the recovered distribution of $H$ magnitudes to match those from the model. 

\medskip

Our estimate of the NEO population agrees most closely with the that of \citet{Stokes2017} for $H<25$ and \citet{Tricarico2017} at smaller sizes. It falls somewhat below the Granvik model and the median of previous observational results, which is perhaps to be expected since most of our potential errors tend toward underestimation. The even lower numbers of small NEOs found by \citet{Trilling2017} may be due in part to their use of the approximation $R^2 \Delta^2 \sim \Delta^4$, where $R$ and $\Delta$ are the heliocentric and geocentric distances to an asteroid, respectively: this approximation is a good one for trans-Neptunian objects, but less so for NEOs. The results of \citet{Lilly2017} and especially \citet{Harris2015} lie well above our estimate for $H>25.0$, though the disagreement is less significant for the very smallest asteroids we have probed. Both of these results appear statistically inconsistent even with our `lost NEO' upper envelope over at least some part of the range faintward of $H=25.0$. They could be brought into agreement by positing ATLAS non-detections of many small NEOs due to large rotational brightness variations --- but these rotationally-caused non-detections would have to outnumber our `lost NEOs' by a factor of a few.

\subsection{Power Law Fits to the NEO Distribution} \label{sec:pwr}

The magnitude and/or size distributions of various populations in the Solar System have frequently been modeled as power laws. Asteroid populations in general (main belt or NEO) do not follow a strict power law over the entire range of observable absolute magnitude and size, but power laws can be a good approximation over significant ranges --- e.g. several magnitudes, or a factor of ten in size. The fits have additional value because the points where they fail (the {\em breaks} in the power laws) are believed to be physically meaningful. For example, \citet{Bottke2005,deElia2007} predict a change in the slope of asteroid size distributions at about 200 meters, because this size marks a transition in physical structure \citep[e.g.,][]{Harris2015}. Larger objects are believed to be strengthless `rubble piles' bound by gravity, while the internal cohesion of smaller objects is due primarily to their nonzero tensile strength, rather than their extremely weak self-gravity.

\medskip

Power laws describing asteroid size distributions can be given in terms of absolute magnitude or size, and can represent either the differential or the cumulative distribution. We use the symbols $\alpha$ and $b$ for the slopes of the absolute magnitude and size distributions, respectively, and subscript $d$ indicating the differential distributions:

\begin{eqnarray} 
  N(H) &\propto & 10^{\alpha_d H}dH \label{eq:pwr01} \\
  N(D) &\propto & D^{-b_d} dD \label{eq:pwr02}
\end{eqnarray}

If the mean albedo of NEOs does not change with size, the magnitude and size slope parameters are related by:

\begin{equation} 
b_d = 5\alpha_d + 1 \label{eq:pwr05}
\end{equation}

The cumulative absolute magnitude distribution is:

\begin{equation} 
  N(<H) \propto 10^{\alpha_c H} \label{eq:pwr03}
\end{equation}

And the differential and cumulative magnitude slopes $\alpha_d$ and $\alpha_c$ are equal, due to the exponential form of Equations \ref{eq:pwr01} and \ref{eq:pwr03}.

\medskip

\begin{figure*}
\plottwo{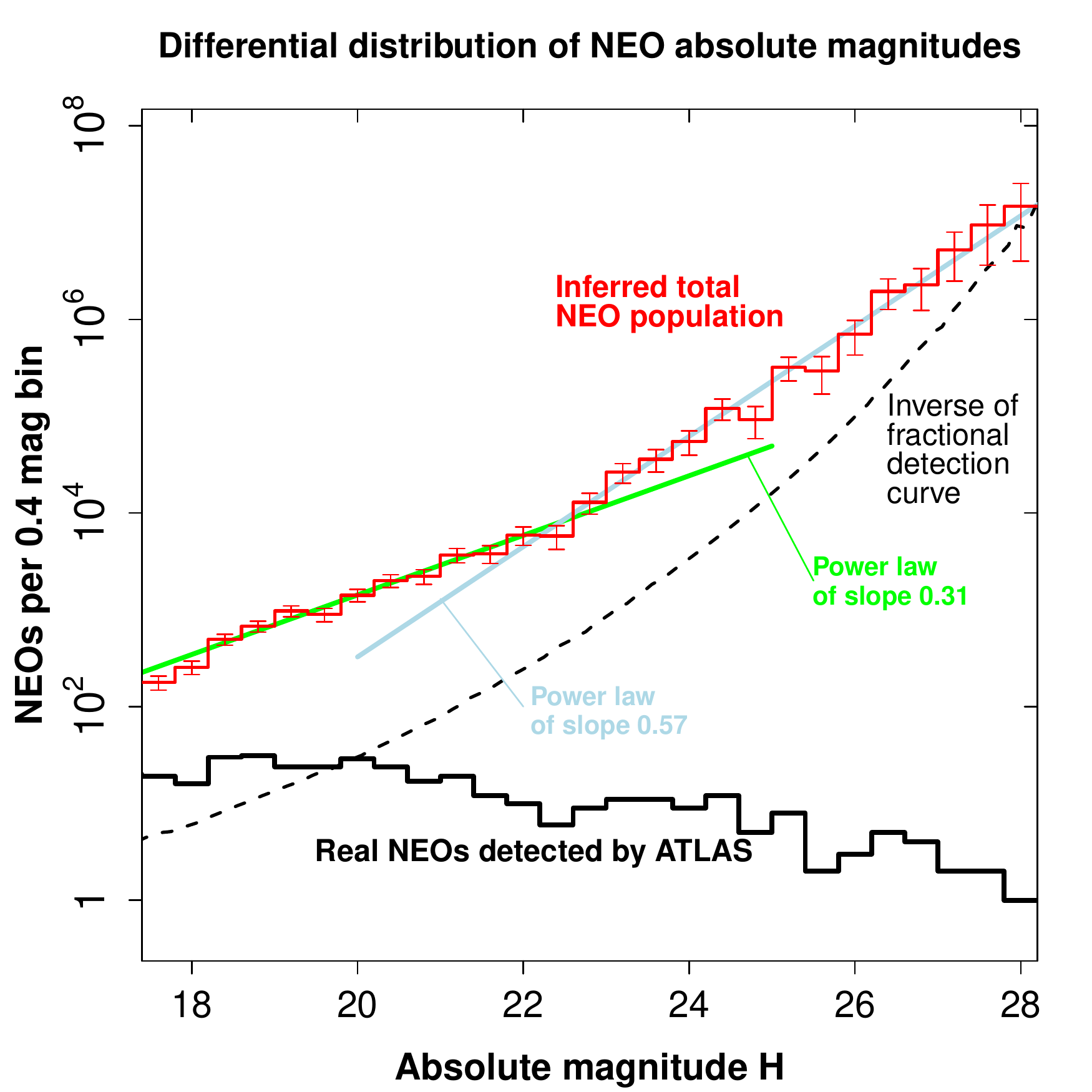}{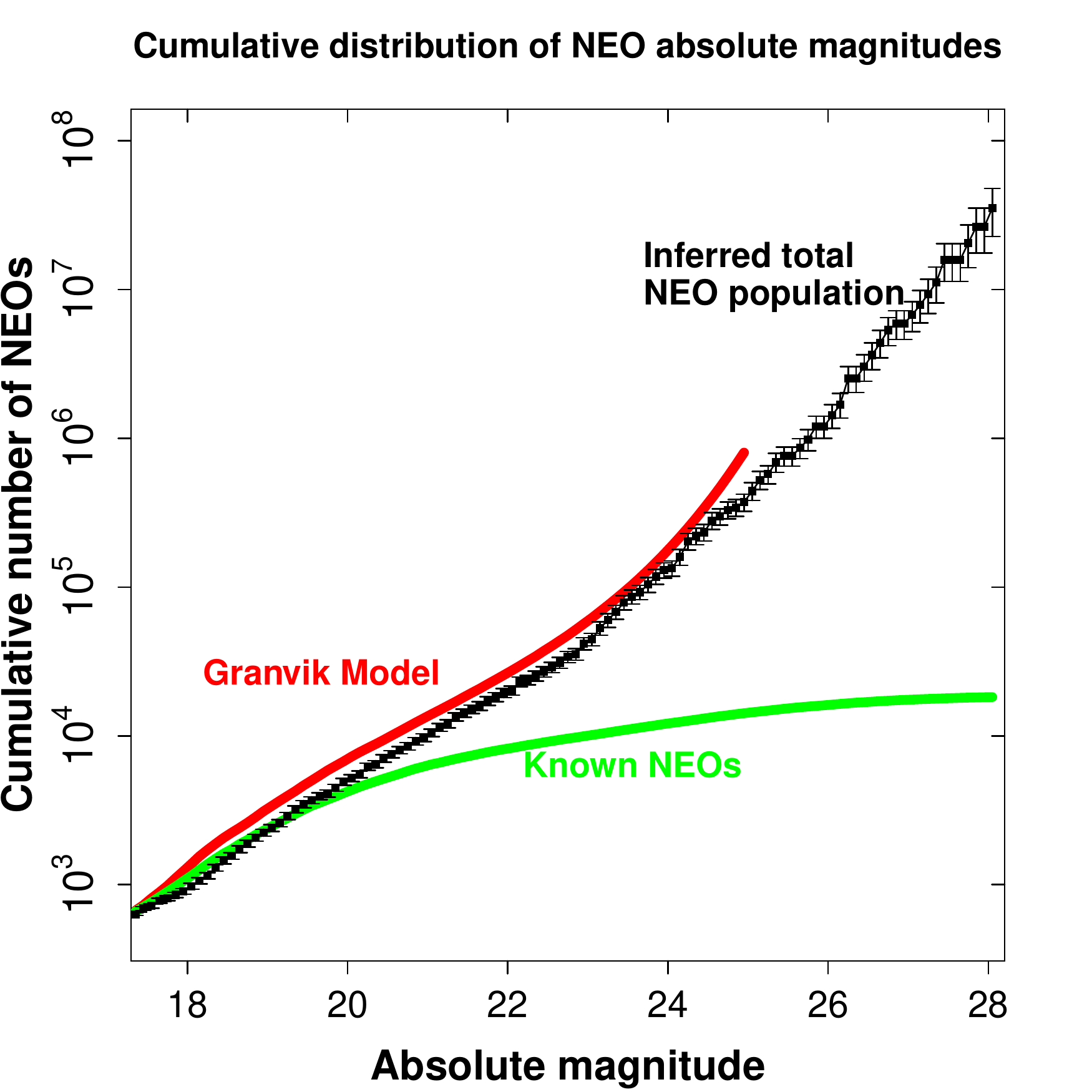}
\caption{{\em Left:} Histogram of NEO absolute magnitudes based on the ATLAS NEO simulation and the 713 distinct real NEOs detected by ATLAS over the same period. The power laws shown are linear least-squares fits to the logarithm of Equation \ref{eq:pwr01} in the ranges $H=$ 18--22 and 23--28, respectively. {\em Right:} The corresponding cumulative distribution from the ATLAS results, compared both to the Granvik model and to the total count of known NEOs.
\label{fig:diffhist}}
\end{figure*}

\begin{figure*}
\includegraphics[width=7.0in]{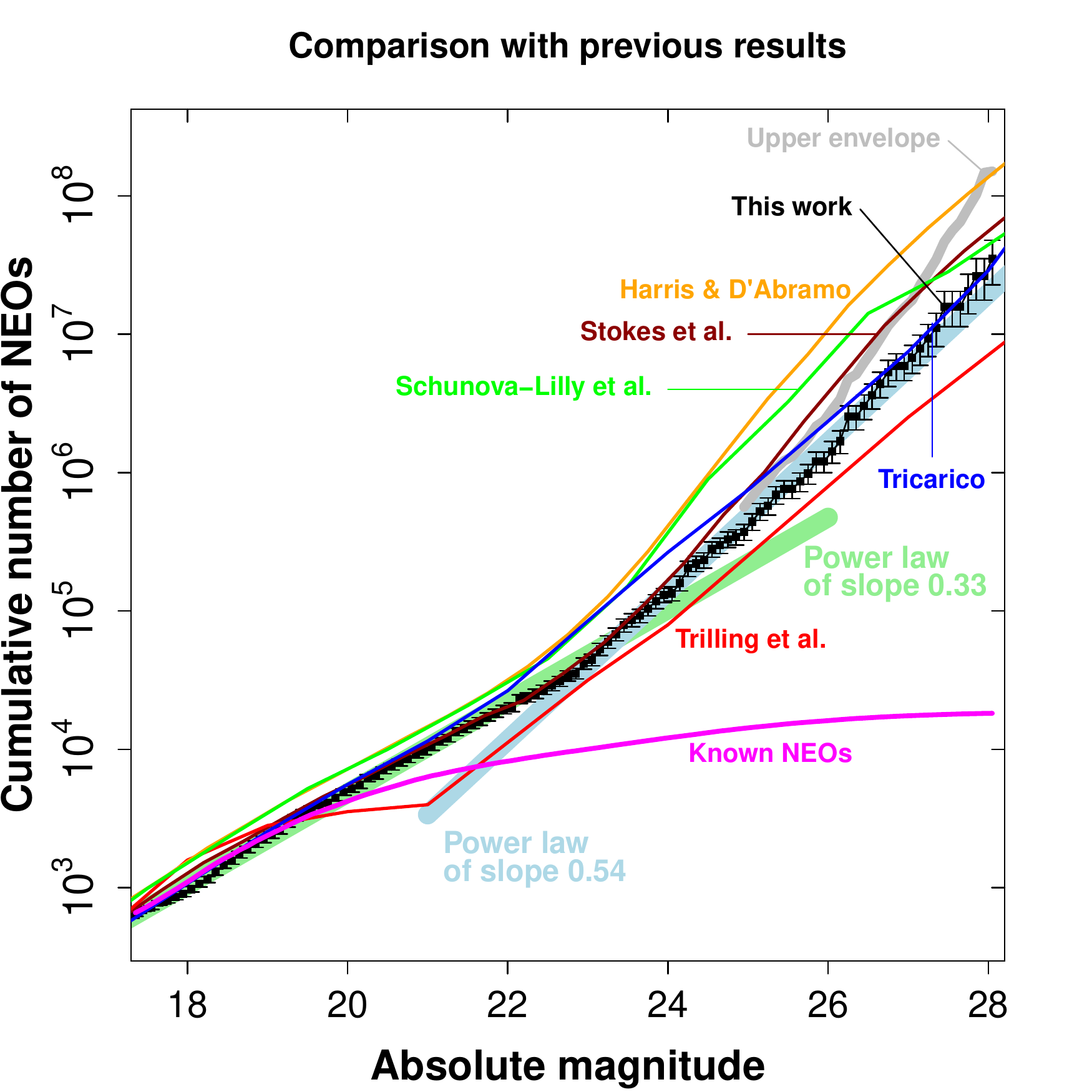}
\caption{Cumulative distribution of NEO absolute magnitudes from ATLAS, compared with power law fits to Equation \ref{eq:pwr03} in the ranges $H=$ 18--22 and 23--28; and with previous results published by \citet{Harris2015,Lilly2017,Stokes2017,Tricarico2017,Trilling2017}. The light gray curve shows the `upper envelope' from ATLAS, as described in Section \ref{sec:lost}.
\label{fig:cumhist}}
\end{figure*}

Figure \ref{fig:diffhist} suggests the corrected histogram $N(H)$ can reasonably be modeled with Equation \ref{eq:pwr01} in two regimes, but that there is a large change in slope between $H=22$ and 23. For $N(H)$ from $H=18$ to 22, we use linear least-squares fitting to the logarithm of Equation \ref{eq:pwr01} to obtain a best-fit value of $\alpha_d = 0.31$, while for $H=23$ to 28, we find $\alpha_d = 0.57$. The fits are formally good, but the slope values should be regarded as very tentative both because of our `lost NEOs' (Section \ref{sec:lost}) and because the large slope change indicates the underlying size distribution is not a perfect power law. Though exact slopes differ, the existence of the slope change near $H \sim 22.5$ has also been indicated by previous work including \citet{Harris2015,Stokes2017}, and the model of \citet{Granvik2018}, and has been linked to the previously mentioned physical transition from larger, strengthless objects to smaller ones with nonzero tensile strength. Our least-squares fits to the cumulative distribution (Equation \ref{eq:pwr03}, again in logarithmic form) find $\alpha_c = 0.33$ and $\alpha_c = 0.54$ over the same respective $H$ ranges (Figure \ref{fig:cumhist}). The reduced difference between the two slopes is likely due to the blurring effect of going from the differential to cumulative distributions.

\medskip

It is interesting to compare the power law slope we find for brighter NEOs with measurements in the main belt --- though such results are tentative because of the difficulty of probing even asteroids as bright as $H=20$ at main belt distances. A slope of $\alpha_d = 0.23 \pm 0.04$, somewhat shallower than our bright-regime value of 0.31, has been measured in the inner main belt for $H=15$ to 19.8 \citep[][Figure 14]{Gladman2009}. Any conclusions from this comparison should be tentative, especially since the \citet{Gladman2009} range only partly overlaps our bright regime of $H=18$ to 22. Since most NEOs originate from the main belt \citep[e.g.][]{Bottke2002}, the slope difference (if real) suggests that the processes transporting asteroids from the main belt into near-Earth orbits are more efficient for smaller objects --- which would not be surprising, since the size-dependent Yarkovsky effect is believed to be important \citep{Farinella1998,Nesvorny2004}.

\medskip

We can convert the absolute magnitude slope values found in our fits to Equation \ref{eq:pwr01} into slopes for the corresponding power law size distribution (Equation \ref{eq:pwr02}), using Equation \ref{eq:pwr05}. We obtain differential size slopes $b_d = 2.5$ and 3.8 for the larger and smaller asteroid regimes respectively. Although these values should be regarded as very tentative for the reasons given above, the differential size slope $b_d = 3.8$ found for the smaller NEOs is interesting because it is close to the boundary value $b_d = 4.0$, beyond which cumulative mass diverges as size goes to zero. This suggests, e.g., that NEOs with sizes from 5--10 meters may contribute nearly as much to the population's total mass budget as objects in the 50--100 meter range.

\medskip

\section{Survey Biases Against High Encounter Velocity} \label{sec:bias}

So far we have characterized the population of NEOs only using a single parameter --- the absolute magnitude $H$ --- and have used the ATLAS NEO simulation for the sole purpose of determining the fractional detection curve $f_d(H)$. Though the simulation was designed to yield $f_d(H)$, it has the potential to reveal other interesting characteristics of the NEO discovery process: for example, the distribution of detected asteroids on the sky, and the relative detectability of similarly-sized asteroids in different orbits. We have probed many such correlations, but herein we report only our most striking result: ATLAS and all other asteroid surveys are almost blind to small ($H>23$) asteroids that encounter Earth at high relative velocity (Figures \ref{fig:velbias01} and \ref{fig:velbias02}). Global discovery statistics for these objects are strongly biased toward NEOs that encounter Earth at relative velocities below the true median.

\subsection{Parametrizing NEO Encounters with the Earth}

We define an encounter between Earth and an asteroid as the moment of closest approach during a given apparition: i.e. any local minimum in the asteroid's geocentric distance as a function of time. The parameter of interest is $v_E$ (the encounter velocity), defined as the asteroid's Earth-relative physical velocity at this moment of closest approach, but we will also use $d_E$, the geocentric distance at close approach, as an important threshold. 

\medskip

Note that $v_E$ and $d_E$ are physical rather than observational parameters. Since they refer to a specific encounter, they are not invariant properties of a given NEO. Instead, an asteroid could have many encounters with Earth and have different values of $v_E$ and $d_E$ each time --- though the characteristics of its orbit would constrain the possible ranges of the encounter parameters. On the other hand, any detection of an NEO (whether simulated or real) can be mapped to unique values of $v_E$ and $d_E$ corresponding to the encounter that is nearest in time to the moment the detection was made. Small NEOs (e.g. $H>23$) are usually only detected fairly close in time to an encounter.

\medskip

\begin{figure*}
\plottwo{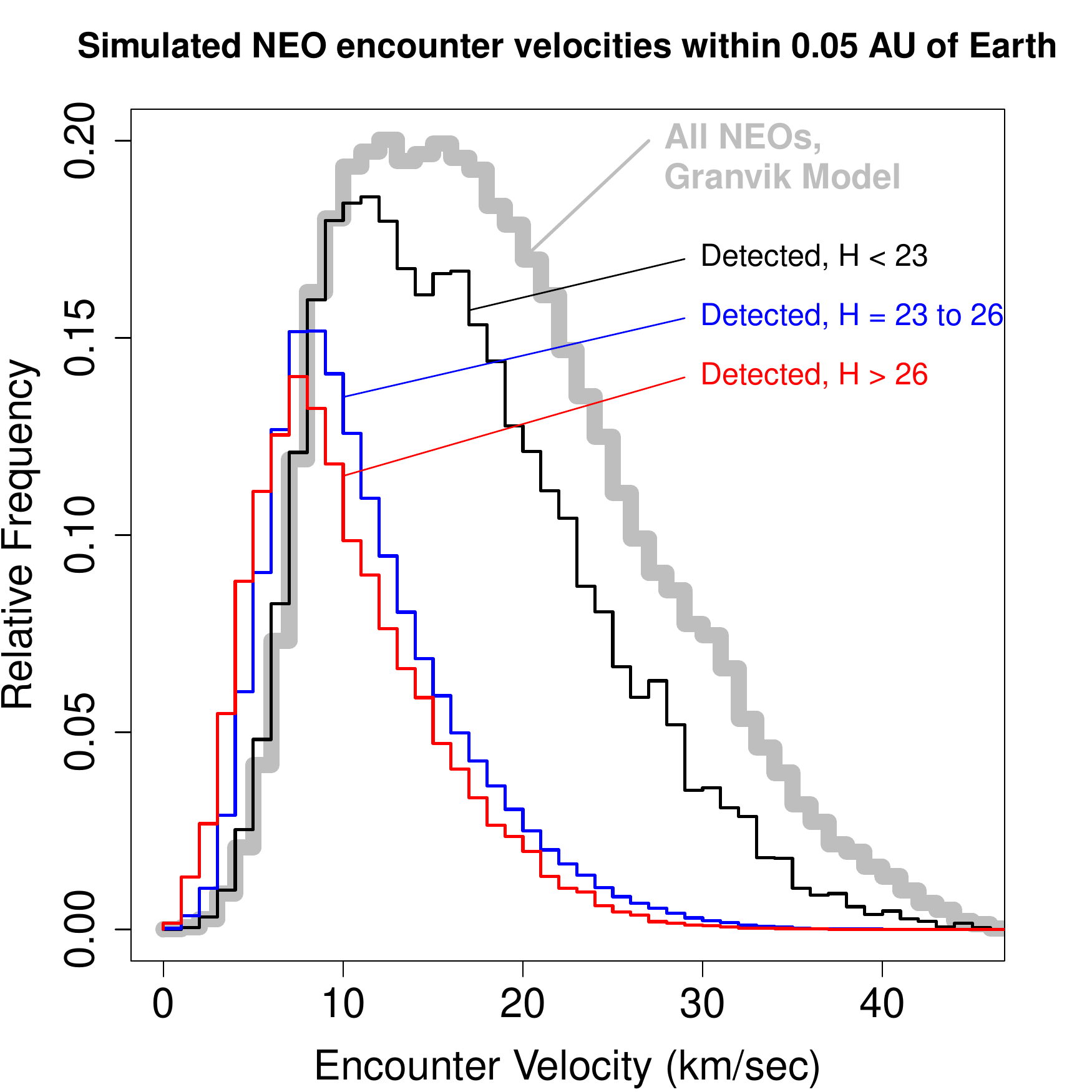}{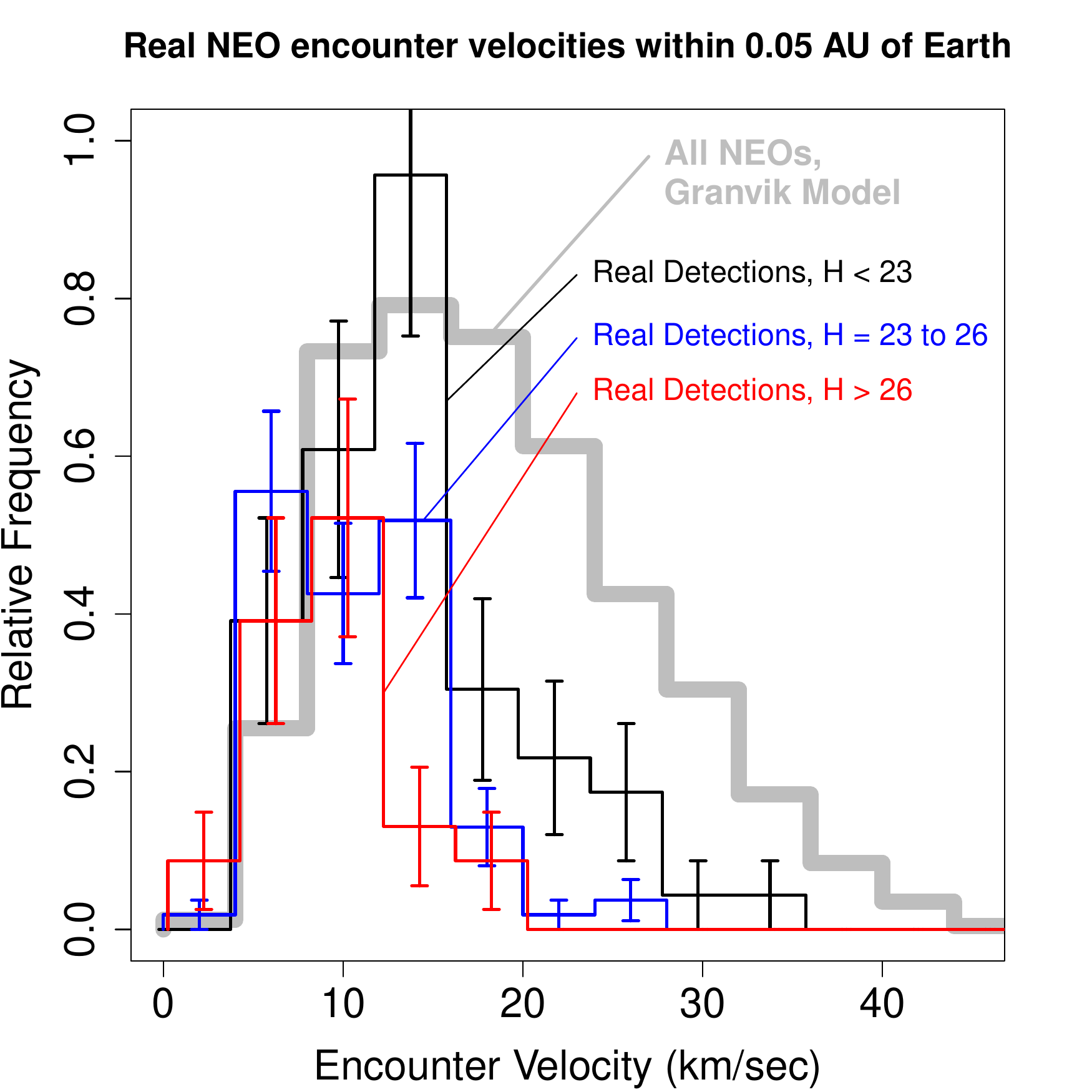}
\caption{Bias against detections/discoveries of small NEOs with high encounter velocities. {\em Left:} ATLAS simulation results. All histograms are normalized to have the same integral from $v_E=0$ to 12 km/sec, so the gap between the Granvik model and the other histograms gives an idea of how many high-velocity objects are missed. {\em Right:} The bias seen in the left panel is not an artifact of our simulation, but also appears in the velocity distributions of real NEOs detected by ATLAS. The histograms have been normalized as at left, but larger bins have been used to reduce statistical noise, and slight horizontal offsets have been added to aid readability. 
\label{fig:velbias01}}
\end{figure*}

We have performed most of our analyses using a maximum encounter distance of 0.05 AU (19.5 times the Earth-Moon distance), since this is close enough that asteroids down to $H=26$ can be detected by most surveys under favorable circumstances. The model of \citet{Granvik2018} allows us to calculate the expected distribution of encounter velocities within this distance. Under the assumptions of our simulation, this distribution is independent of absolute magnitude. We show it as a gray histogram in Figure \ref{fig:velbias01}. 

\subsection{A Worldwide Bias}

In our simulation, the distribution of encounter velocities for {\em detected} NEOs is substantially weighted toward low velocities relative to the `true' distribution from the Granvik model. The bias exists for NEOs with $H<23$, but becomes much stronger for smaller objects with $H=23$ to 26, and stronger still at $H>26$ (Figure \ref{fig:velbias01}, left panel). The simulation is telling us that the ability of ATLAS to detect small NEOs is greatly reduced if they have high encounter velocities. As illustrated by the right panel of Figure \ref{fig:velbias01}, the encounter velocity histograms of {\em real} NEOs detected by ATLAS during the period of the simulation indicate the same thing, though with more noise due to small-number statistics.

\begin{figure}
\includegraphics[width=3.5in]{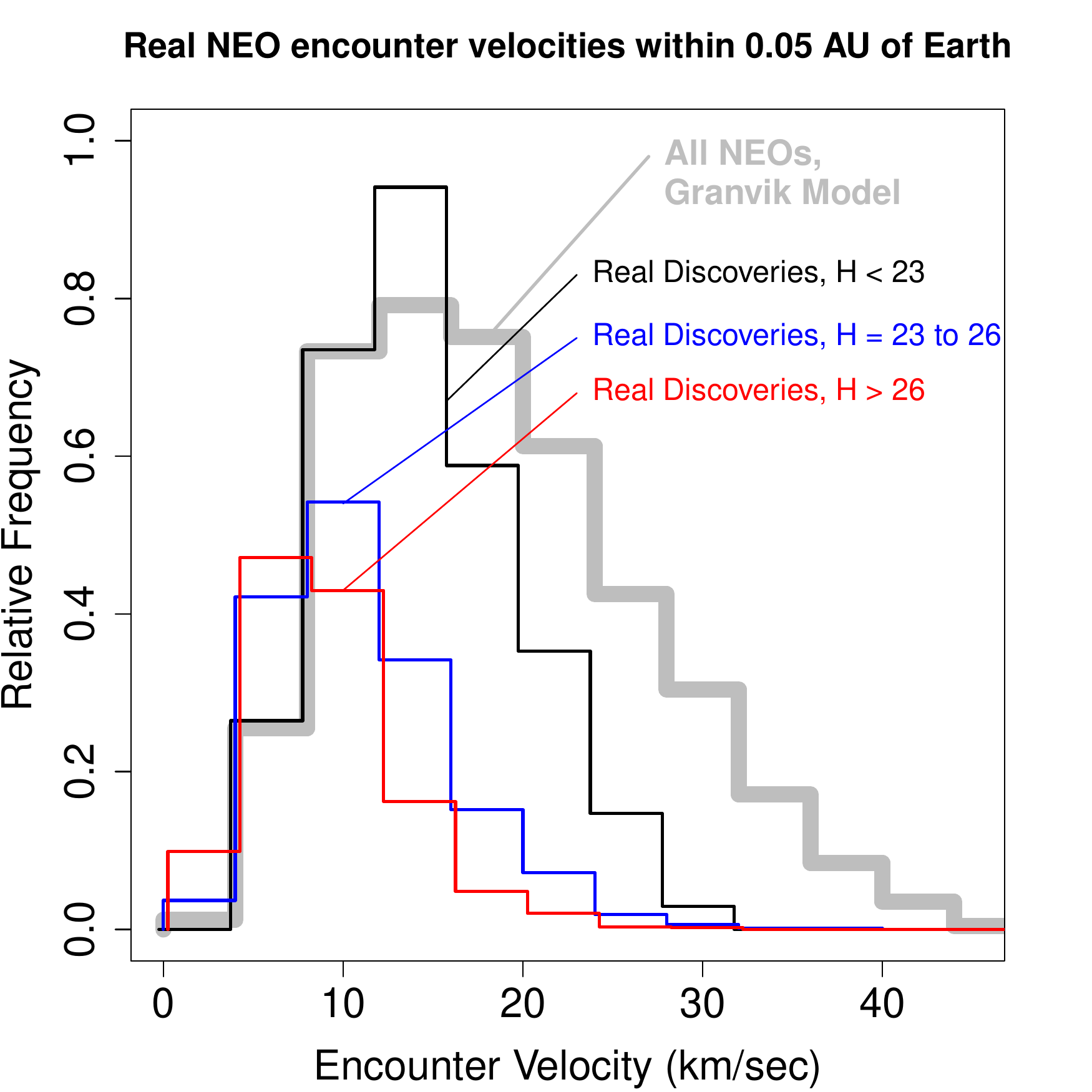}
\caption{The bias seen in Figure \ref{fig:velbias01} is not confined to the ATLAS survey. Here we show histograms of real NEO discoveries for all observatories worldwide between 2016 January 01 and 2019 August 22, normalized for $v_E \leq 12$ km/sec as in the previous figure. The bias against detection of small asteroids with fast encounter velocities persists.
\label{fig:velbias02}}
\end{figure}

\medskip

To determine if the bias against high encounter velocities is unique to ATLAS, we obtained encounter parameters for all NEOs discovered worldwide between 2016 January 01 and 2019 August 22. We found that over this period, 2827 NEOs were discovered in the course of an Earth encounter that took them inside our 0.05 AU limit. We did not include NEOs discovered earlier and recovered in the course of close encounters during this period, which would have required a deeper dive into worldwide asteroid detection statistics. This will have little effect in the size range of interest, however: only a small fraction of NEOs with $H>23$ are known, so new discoveries greatly outnumber objects that are returning after being discovered in a previous apparition. Figure \ref{fig:velbias02} shows the encounter velocity histograms of these 2827 real NEOs, divided into the same ranges of absolute magnitude as we used for the ATLAS results in Figure \ref{fig:velbias01}. The bias against high encounter velocities for worldwide NEO discoveries is at least as strong as that seen in the ATLAS simulation.

\medskip

To quantify this in more detail, we consider the 1259 real NEOs plotted in Figure \ref{fig:velbias02} for the $H=23$ to 26 range. This range can be considered as the smallest cohort of NEOs whose impacts remain dangerous --- hence, they are strongly affected by encounter velocity bias and yet are large enough to be important for planetary defense. The true median and 3rd-quartile encounter velocities for objects passing within 0.05 AU of Earth are 17.5 and 23.9 km/sec, respectively, from our simulations based on the Granvik model. Of the 1259 real NEOs with $H=23$ to 26 discovered during encounters at least this close, only 131 (10\%) had encounter velocities faster than 17.5 km/sec and only 25 (2\%) were faster than 23.9 km/sec. For an unbiased survey, of course, these fractions would be 50\% and 25\%. If we take the 1128 NEOs discovered with encounter velocities below the 17.5 km/sec median as a guide to the number of asteroids that {\em actually} passed Earth with above-median speeds, the implication is that we are blind to 88\% of objects with encounter velocities above the median and 96\% of those with velocities above the third quartile. Unfortunately, these faster objects contribute disproportionately to the planetary risk, as we discuss below.

\begin{figure*}
\plottwo{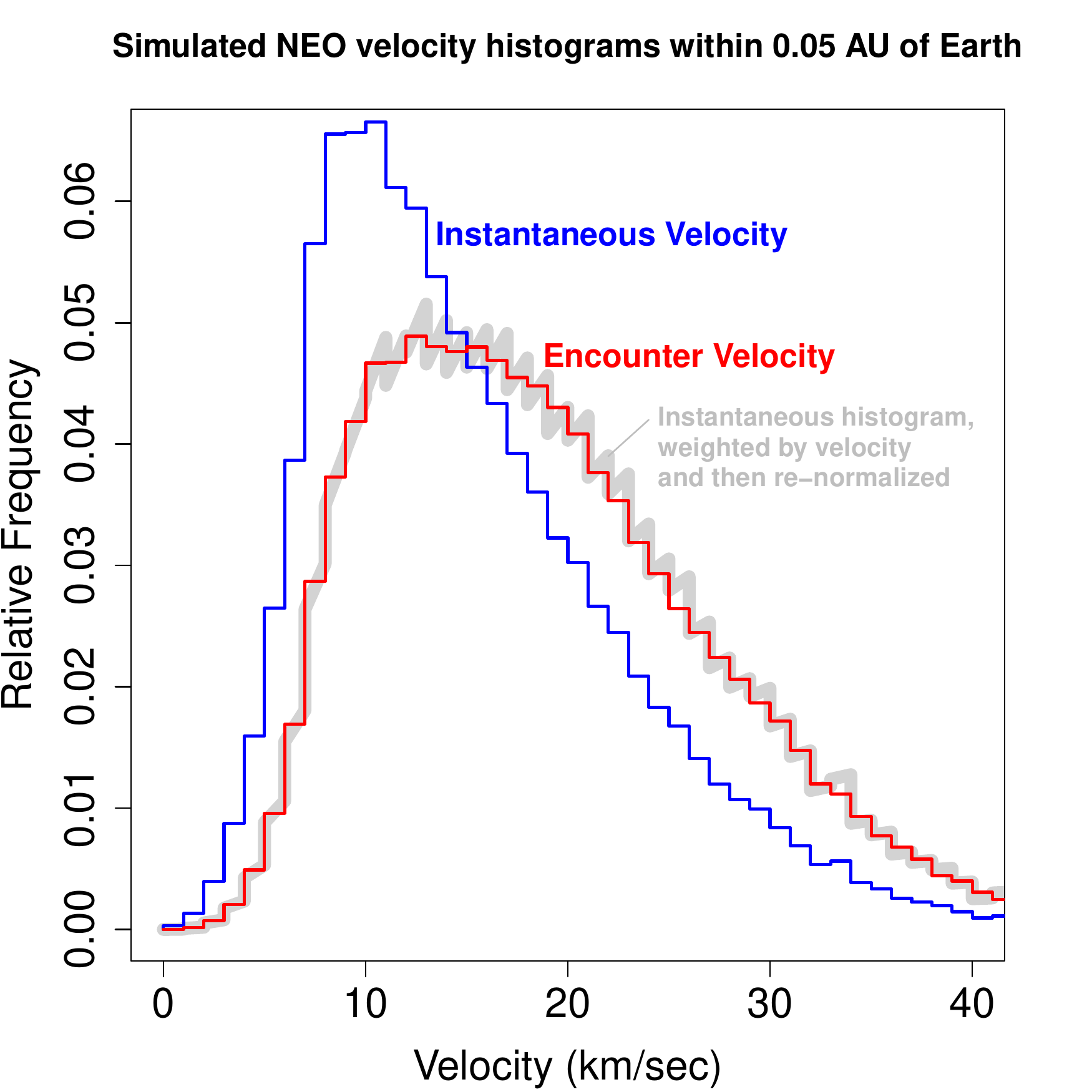}{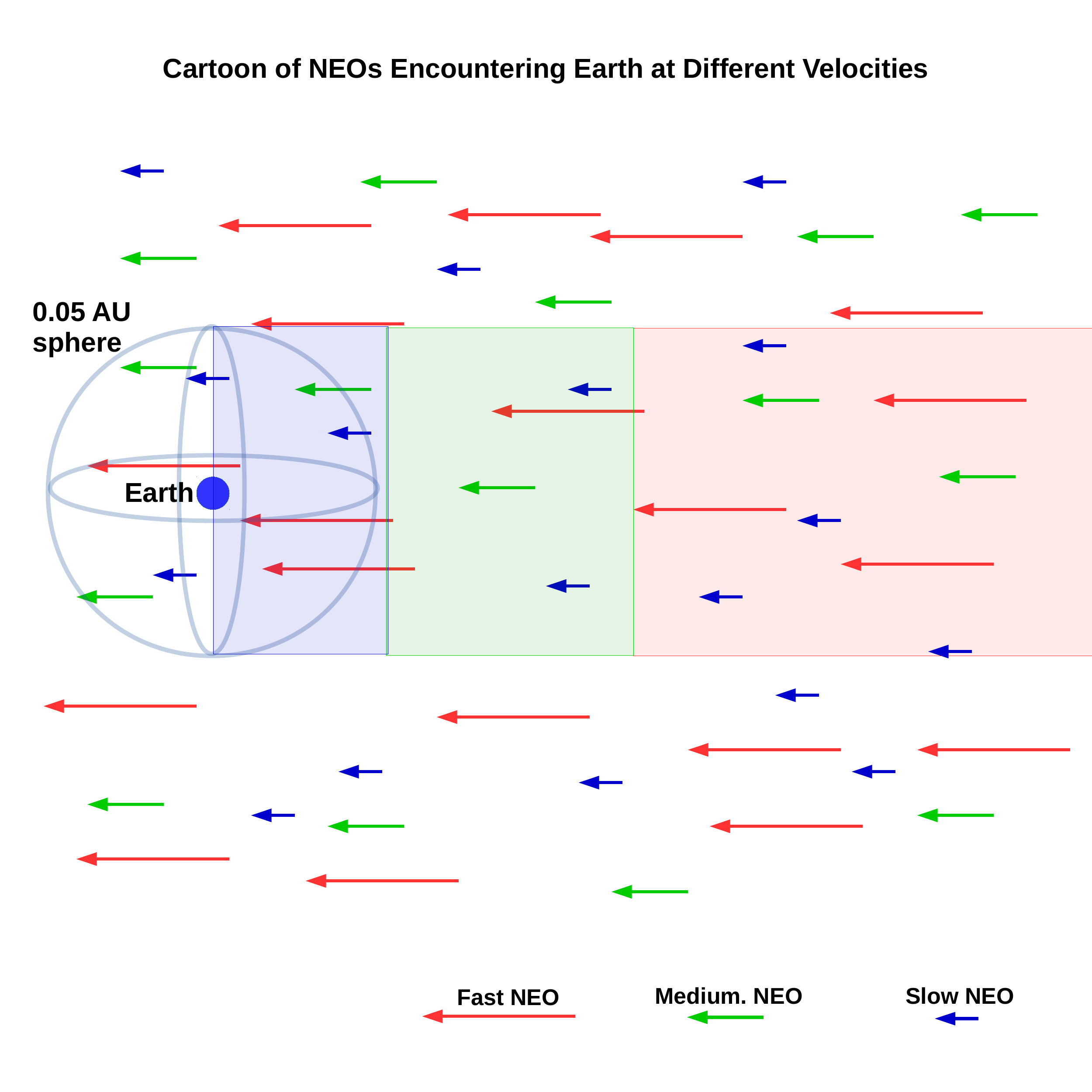}
\caption{The distribution of Earth-encounter velocities for NEOs is weighted toward faster speeds relative to the instantaneous distribution of NEO velocities in near-Earth space, because fast-moving asteroids sweep out more volume of space per unit time and are therefore more likely to fly past Earth. {\em Left:} Both distributions from a simulation using the NEO orbital distribution from \citet{Granvik2018}, showing how the encounter velocity curve can be recovered by velocity-weighting the instantaneous distribution. {\em Right:} Cartoon illustrating why this works. Only asteroids within the sphere described around the Earth contribute to the instantaneous distribution, with all speeds equally represented in the cartoon. By contrast, the volume of space inhabited by asteroids that will {\em encounter} Earth within a given time span is smaller (blue-shaded volume only) for slow NEOs and larger (all the shaded volumes together) for fast ones. All velocities considered here are relative to Earth.
\label{fig:instant01}}
\end{figure*}

\subsection{Instantaneous Velocities vs. Encounter Velocities}

It is important to note that the distribution of true encounter velocities plotted in Figure \ref{fig:velbias01} is {\em not} the same as the instantaneous distribution of physical relative velocities for NEOs within 0.05 AU of Earth at a given moment. Both of these distributions can be calculated from our simulation based on the Granvik model, and the distribution of encounter velocities has relatively more weight at faster speeds. To see why, consider how both distributions can be extracted from a simulation. The instantaneous velocity distribution is found by `freezing' the Solar System at a particular moment, counting up all the NEOs within 0.05 AU of Earth, and calculating the histogram of their physical velocities relative to Earth\footnote{In practice, for a finite-sized simulation, it may be necessary to perform this `freeze and count' process many times in order to build up sufficient statistics. We spaced these `freeze times' widely enough to avoid double-counting of slow asteroids, though this turns out not to affect the result.}. By contrast, the distribution of encounter velocities is calculated over time, where each NEO passing within 0.05 AU of Earth contributes a single point --- its velocity at closest approach --- to the histogram.

\medskip 

Another way to think about this is in terms of individual NEO orbits. We can classify an orbit as dangerous if its MOID (minimum orbital intersection distance) relative to Earth is less than 0.05 AU. But encounters with $d_E<0.05$ AU are not equally frequent for all such NEOs. In general, we expect the danger from NEOs in a given orbit to scale with the frequency of close encounters. We can think of the distribution of true encounter velocities (gray histograms in Figures \ref{fig:velbias01} and \ref{fig:velbias02}) as the sum of the encounter velocity distributions for individual orbits, weighted by the frequency of close encounters for those orbits. The instantaneous velocity distribution is, likewise, the sum of individual-orbit velocity distributions weighted by the frequency of close encounters --- but it is weighted additionally by the typical {\em duration} of encounters for a particular orbit: i.e., by the `dwell time' an NEO spends near Earth during an encounter. This `dwell time' is inversely proportional to encounter velocity: hence, as illustrated by Figure \ref{fig:instant01}, the distribution of instantaneous relative velocities of NEOs near Earth is weighted toward slower speeds relative to the distribution of encounter velocities.

\subsection{Encounter Velocities and Orbits}

\begin{figure}
\includegraphics[width=3.5in]{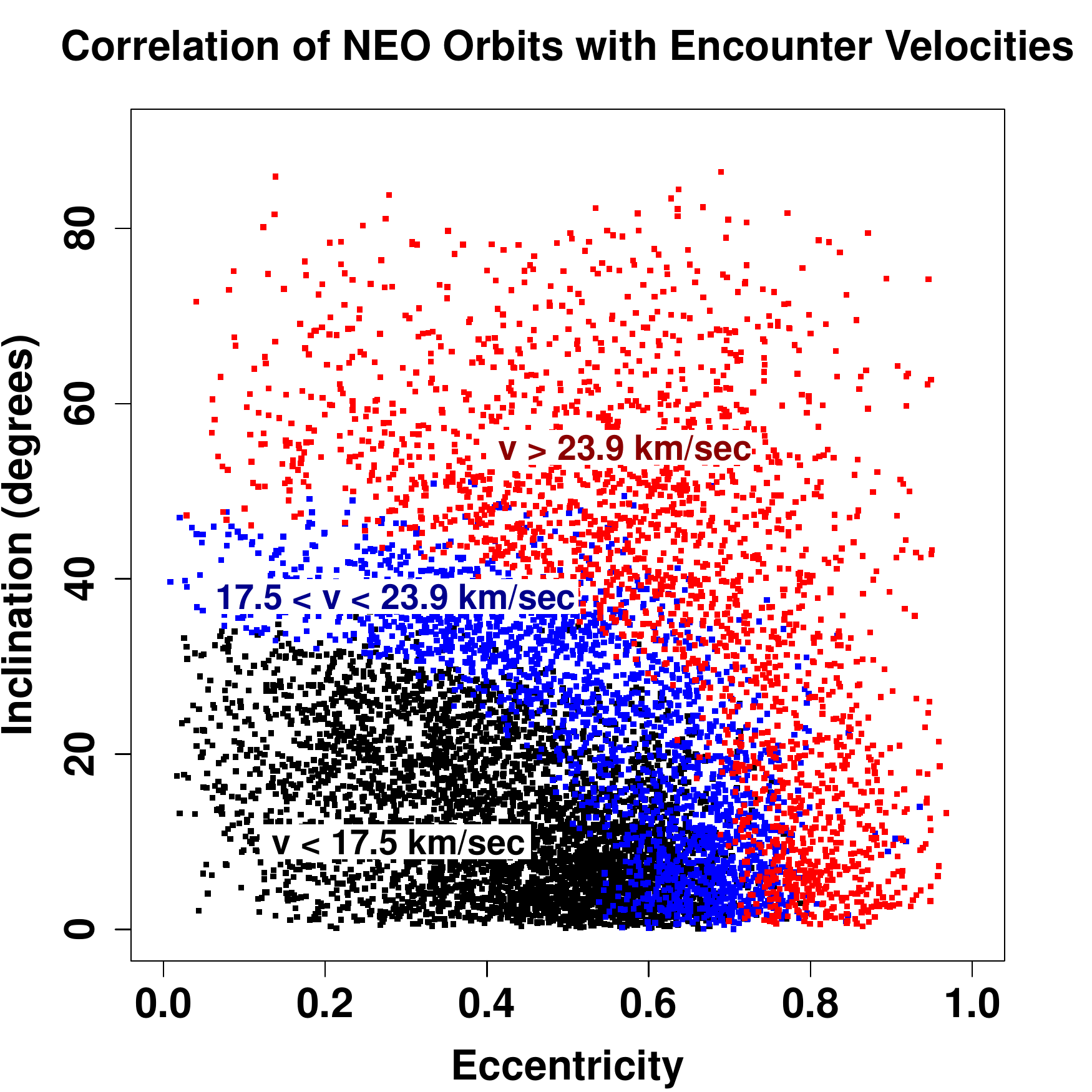}
\caption{Relative velocities of NEOs encountering Earth depend on the asteroids' orbital eccentricity and inclination, and orbits exist that never encounter Earth at low velocity. In these results from the ATLAS simulation, orbits corresponding to NEOs with Earth encounters below the median of 17.5 km/sec are plotted in black; those with encounter velocities between the median and third quartile (23.9 km/sec) in blue; and the fastest quartile is plotted in red. Except for very large NEOs, current surveys are nearly blind to the asteroids in blue and (especially) red. Hence, there are certain orbits we cannot effectively probe.
\label{fig:orb}}
\end{figure}

The bias against finding NEOs with high encounter velocities makes current surveys effectively blind to small NEOs in certain types of orbits. As illustrated by Figure \ref{fig:orb}, NEOs in orbits with high eccentricity or (even more definitively) high inclinations {\em cannot} encounter Earth at low velocities. Hence, we have almost no observational ability to constrain the abundance of NEOs in such orbits (except for very large objects). Note that this orbital bias applies {\em within} the set of NEOs that actually have close encounters with Earth, and exists independent of (and in addition to) the well-known fact that orbits with high eccentricity and/or inclination are less likely to enable close approaches to Earth. 

\medskip

In principle, the simulation we have described herein would be capable of debiasing ATLAS detections of NEOs in orbits with high eccentricity and/or inclination --- that is, orbits that inevitably produce high encounter velocities. We could do this by adding additional terms to the fractional detection function $f_d(H)$ --- e.g. $f_d(H,v_E)$ or $f_d(H,e,i)$ with $e$ and $i$ the orbital eccentricity and inclination, respectively. This will not work in practice, however, because not enough real NEOs in such orbits were detected by ATLAS: the statistics are too noisy. Of the 713 real NEO detections that make up our $D(H)$ distribution (Equation \ref{eq:diff}), only four had $H \geq 23$, $d_E \leq 0.05$ AU, and encounter velocity above the model median of 17.5 km/sec. Worse still, it is quite likely that some of the 13 lost asteroids belonged in this category: high encounter velocity is correlated with high angular velocity, which in turn makes NEO candidates more likely to go unconfirmed \citep{Veres2018}.

\medskip

The worldwide catalog of 2827 NEO discoveries we used to make Figure \ref{fig:velbias02} includes 131 objects with $23 \leq H \leq 26$ and encounter velocities greater than the 17.5 km/sec median. Of these, 25 had $v_E$ greater than the 23.9 km/sec third quartile. These numbers could offer sufficient statistics for interesting constraints on the population of small NEOs in high-velocity orbits --- except, of course, that the total performance of worldwide NEO discovery capability cannot be simulated at the `Solar System to pixels' level we have used to derive the fractional detection curve for ATLAS. Also, the imperfect fidelity of any simulation and the hard-to-quantify errors from lost NEOs become most serious exactly in regimes where the detection fraction is low and the probability of candidate NEOs going unconfirmed is high.

\medskip

Hence, the population of $H \geq 23$ NEOs in orbits with high eccentricity and/or inclination could likely differ from theoretical expectations by a factor of a few, without this fact being apparent from current survey statistics. This reality is the reason for our claim in Section \ref{sec:simdetail} that there is (for the present) no better option for a simulation such as our own than adopting our approximation that the orbital distribution of small NEOs is the same as that of larger ones. We have argued tentatively on theoretical grounds that the distributions are probably not greatly different, and although small differences almost certainly exist, current observational data do not enable us to quantify them. This need not always be the case, but careful design both of simulations and survey strategies, as well as efforts to minimize the loss of NEO candidates, will be needed to map the orbital distributions of small NEOs with confidence. Part of our motivation in writing this paper is to inspire such efforts.

\medskip

In the future, one aspect of such work will be leveraging the capability of the Vera C. Rubin Observatory (formerly the Large Synoptic Survey Telescope), which is optimized to address cosmological questions, but nevertheless will detect fainter NEOs than any currently operating survey. It is intended to scan the accessible sky once every three nights (in a variety of filters with greatly differing sensitivity to asteroids), and to take only two images per field per night \citep{LSST}. Despite its impressive capabilities, the Rubin Observatory by itself probably will not produce orbits for large numbers of small, fast NEOs. They are exactly the objects whose brief observing windows and rapid motions across the sky, combined with the observatory's relatively sparse cadence, are most likely to prevent detection and linkage across multiple nights. The new observatory's productivity for such objects will depend on external followup --- and on the invaluable, faint precoveries and recoveries it will produce for NEOs discovered elsewhere. Followup will be difficult because most Rubin Observatory detections will be so faint, and because ephemeris predictions based on only two measurements will have greater uncertainty than those from current surveys using three or more images. Hence, the new observatory's greatest contribution for small, high velocity NEOs will probably be the fact that seemingly lost NEO candidates from other surveys are likely to have faint (p)recoveries in Rubin Observatory data, which, once identified, will instantly enable the calculation of accurate orbits and $H$ mags. Developing techniques to mine Rubin Observatory data for such (p)recoveries is likely to be a worthwhile enterprise.

\section{Velocity Bias and Impact Risk} \label{sec:biasrisk}

\subsection{Causes of the Velocity Bias}

The existence of a strong bias against detection of NEOs with high encounter velocities should not come as a surprise. Several known effects conspire to make the detection of such asteroids more difficult. Some of the most significant of these effects are direct results of the high angular velocities that such objects exhibit: that is, how rapidly they move across the sky.

\medskip

Consider two NEOs of the same size passing Earth with the same encounter distance $d_E$, but with encounter velocities $v_E$ differing by a factor of two. At every stage of the encounter, the angular velocity of the faster asteroid will be twice as large. The high angular velocity makes the asteroid's trailed image longer, fainter, and harder to detect. If detected at all, the faster asteroid's high angular velocity means the uncertainty of its on-sky position will grow rapidly, meaning it is likely to be lost unless followup observations are obtained very promptly. An example of this latter effect is seen in the finding by \citet{Veres2018} that the angular velocities of unconfirmed (i.e., `lost') NEO candidates skew higher relative to those of successful discoveries. 

\medskip

Besides the direct effects of high angular velocity, there are other correlated effects. The faster asteroid's geocentric distance will change more quickly, causing linear acceleration of its on-sky trajectory, which may inhibit tracklet-linking software that starts from the approximation of constant-velocity Great Circle motion. Finally, the higher velocity asteroid's total `dwell time' near Earth is short, meaning that weather, equipment problems, or simply a failure to survey the right part of the sky at the right moment can prevent its discovery.

\subsection{Very Close Encounters and Impacts} \label{sec:vclose}

The direct effects of fast angular velocity are mitigated if a high-speed asteroid is making a {\em very} close approach to the Earth ($d_E << 0.01$ AU). In this case, it approaches the Earth along a nearly radial trajectory, so its angular velocity remains low except for a brief period near the moment of greatest proximity. 

\medskip

Such an asteroid will also become unusually bright for its size, since it approaches so close to the Earth. This is true almost regardless of the encounter geometry, since an asteroid that comes from the direction of the Sun (and hence is invisible while it is approaching Earth) will become bright and easy to see as soon at it has {\em passed by} and begun receding from Earth. Hence, most asteroids with very close encounters give us a good chance to discover them either incoming or outgoing --- with the exception, of course, of objects that impact Earth and have no outgoing trajectory. Such objects cannot be detected by ground-based surveys if they approach from the direction of the Sun \citep[e.g., the Chelyabinsk impactor; see][]{Dunham2013}.

\medskip

Unlike the direct effects of angular velocity, the other effects that tend to prevent the detection of asteroids with high encounter velocities --- angular acceleration and short time spent near Earth --- are {\em not} necessarily reduced in the case of extremely close approaches and impacts. Hence, it is possible that the velocity bias we have noted applies at a reduced level even to objects incoming for impact, making it a significant cause of concern for planetary defense. 

\subsection{Angular Velocity Distributions} \label{sec:angveldist}

We can combine the orbital distribution from the Granvik model with the power law fits from Figure \ref{fig:diffhist} to predict the angular velocity distributions of NEOs observable from Earth. Comparing such a prediction with observed distributions can illuminate the question of which effects contribute most to the velocity bias. To make this comparison, we perform a new simulation spanning one year for a Solar System overpopulated by a factor of thirty relative to our broken power law model of the real $H$ mag distribution ( Figure \ref{fig:diffhist}). This new simulation does not involve painting simulated images into actual data, and hence is not a `Solar System to pixels' simulation. Instead, it merely produces ephemerides at 1-hour sampling for all of the simulated asteroids that reached apparent magnitudes of 20.0 or brighter, at solar elongation greater than $45^{\circ}$ and declination north of $-50^{\circ}$. 

\medskip

To produce a meaningful angular velocity distribution from these ephemerides, we select the angular velocity corresponding to the moment when each simulated asteroid was easiest to detect. For an asteroid with angular velocity faster than 1.6 deg/day (corresponding to a 2-arcsecond trail on a 30-second exposure) we define ease of detection not by the total brightness, but by a `trail intensity' equal to the brightness per unit 2-arcsecond segment of the asteroid's trailed image. An asteroid moving at 16 deg/day leaves a 20 arcsecond trail and has a trail intensity a factor of 10 (2.5 magnitudes) fainter than its total brightness. 

\medskip

Note that our adoption of trail intensity, which decreases linearly with angular velocity, as our metric for detectability represents a conservative evaluation of asteroid surveys' ability to detect trailed objects. The metric makes an implicit assumption that the `trailing loss' (that is, the factor by which the faint-flux limit of a survey's sensitivity deteriorates for trailed images) also goes linearly with trail length. With an optimized matched filter (i.e., cross-correlating the image with a kernal representing a trailed asteroid image), one can in principle achieve a sensitivity that decreases only as the square root of the angular velocity. Efforts to improve sensitivity to trailed images are underway by other surveys \citep[e.g.][]{Ye2019} and by ATLAS, which already achieves slower-than-linearly decreasing sensitivity in the 4-12 deg/day angular velocity regime, reaching trail intensities fainter than 20.5 mag near 10 deg/day despite its 19.5 mag limit for stationary sources. However, these capabilities are not yet sufficiently advanced nor deployed broadly enough to make our linearly decreasing trail intensity an overly pessimistic metric --- as will be seen below.

\medskip

\begin{figure*}
\plottwo{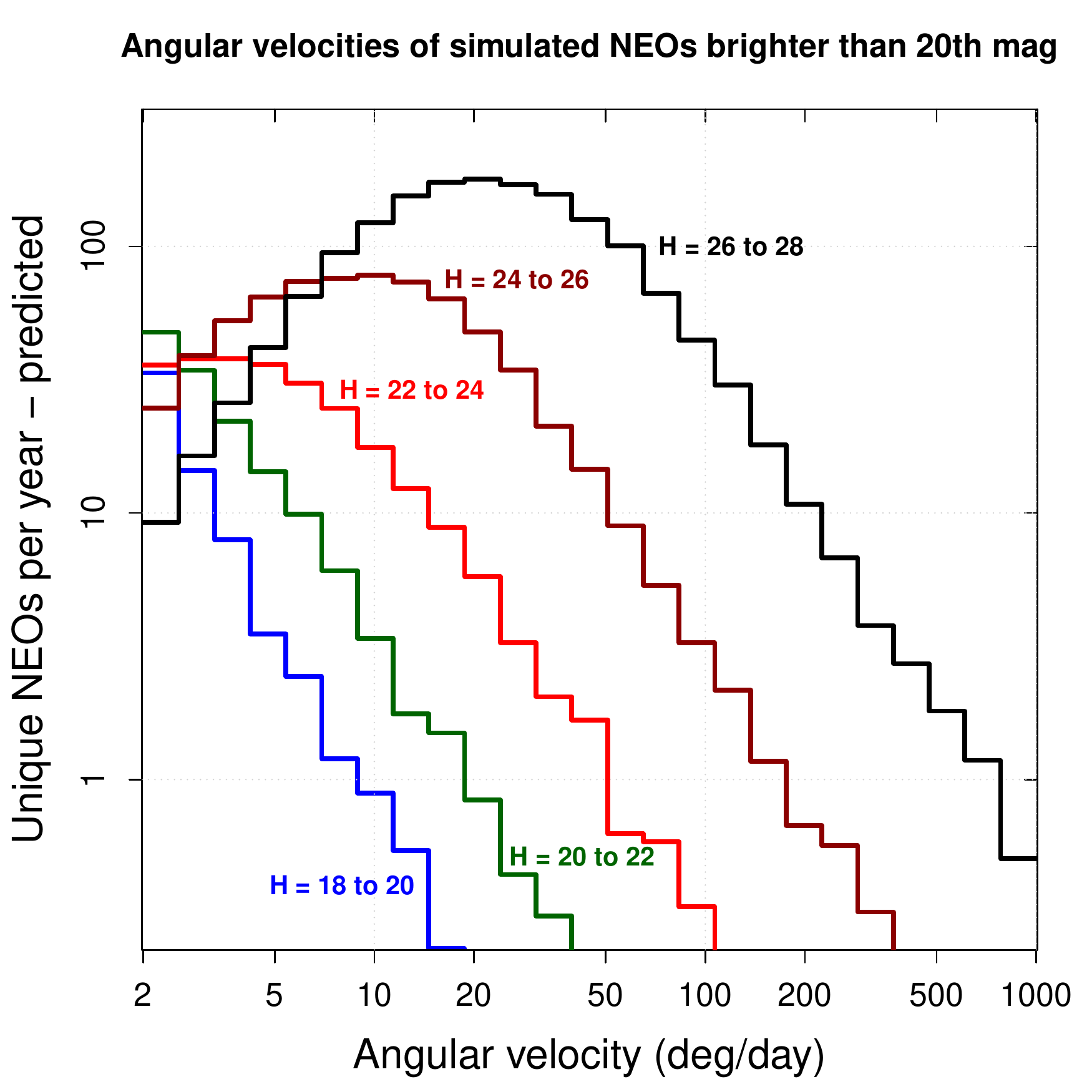}{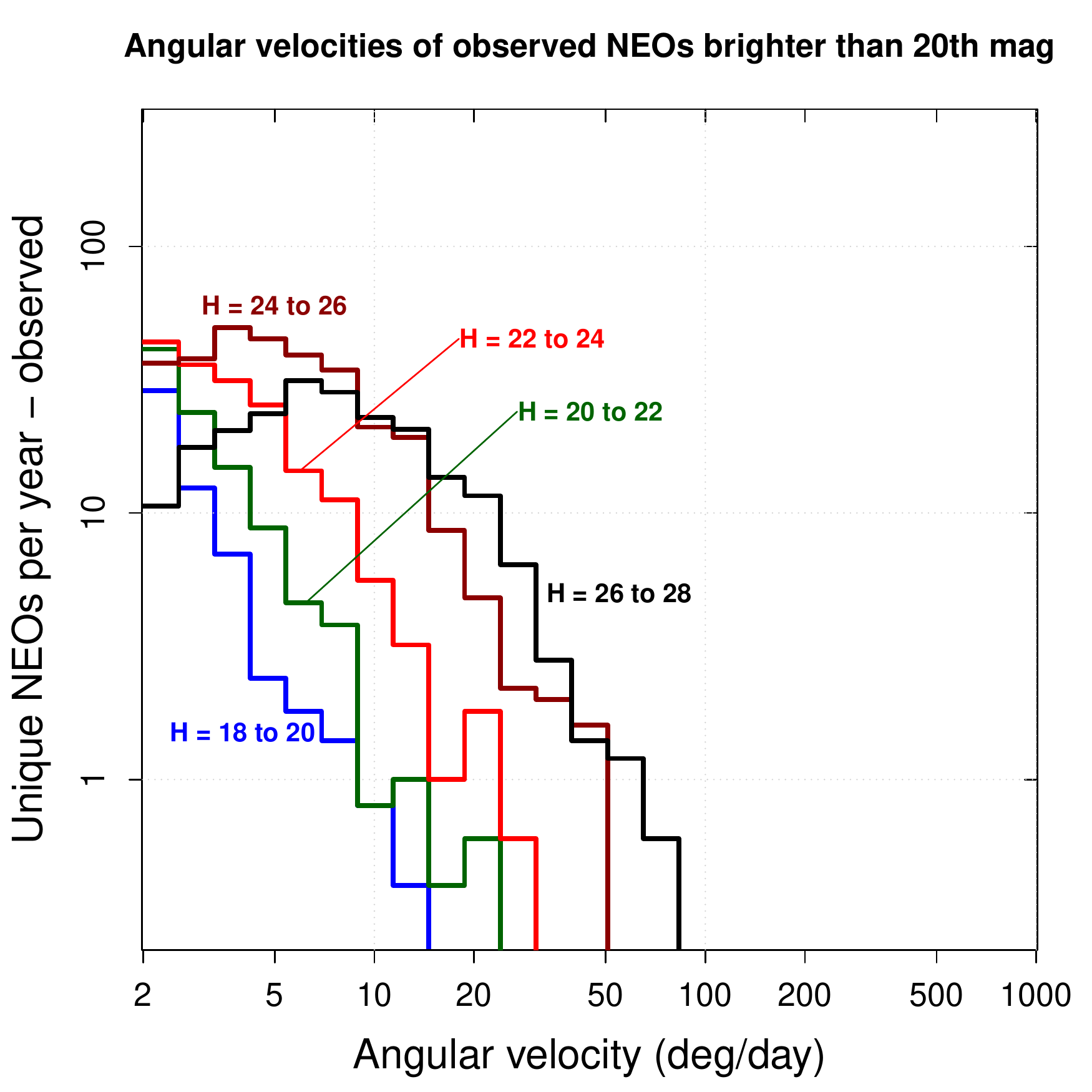}
\caption{Predicted vs. observed angular velocity distributions over one year for NEOs observable from Earth at 20th mag or brighter. Each distinct NEO is counted only once, at the angular velocity corresponding to its maximum observability, defined by maximum brightness per unit 2 arcsecond length of its trailed image on on a hypothetical 30 s exposure tracked at sidereal rates. {\em Left:} Simulated results. Small asteroids tend to have faster angular velocities because they are detectable only when close to Earth {\em Right:} Observed results worldwide for the years 2015 through 2019. Fewer NEOs are observed at high angular velocity relative to the prediction. This bias is negligible for the largest asteroids, but can exceed a factor of 100 for $H>22$ and angular velocity greater than 50 deg/day.
\label{fig:angvelpred01}}
\end{figure*}

Having determined the angular velocity of each simulated asteroid at its moment of brightest trail intensity, we divide the asteroids into bins of $H$ magnitude, and plot the resulting angular velocity histograms in the left panel of Figure \ref{fig:angvelpred01}. These histograms have been normalized by dividing out the $30\times$ overpopulation factor used in our simulation, and hence they constitute an actual prediction of the number of distinct NEOs observable at apparent mag 20 or brighter in a given year, in the respective ranges of $H$ mag and angular velocity.

\medskip

In order to compare these angular velocity distributions with real data, we obtained from the Minor Planet Center all the records of NEOs observed from any observatory in the five-year period from 2015--2019, inclusive. We divided these records into `tracklets' each corresponding to the set of observations of one NEO from one site on a single night. For each tracklet, we calculated the angular velocity, mean magnitude, and trail intensity (without attempting color corrections for the heterogeneous sets of filters used by observatories around the world). This enabled us to select the tracklet corresponding to maximum trail intensity for each real NEO, just as we had for the simulated objects. The histograms of the corresponding angular velocities are plotted in the right panel of Figure \ref{fig:angvelpred01}. 

\medskip

Figure \ref{fig:angvelpred01} dramatically illustrates our worldwide inability to detect a large fraction of the NEOs with high angular velocities --- a shortfall which especially affects the faintest three bins, at $H \ge 22$. The figure indicates that dozens to hundreds of these objects brighten past 20th mag every year without being detected by any telescope, and the detection rate drops below 1\% at angular velocities beyond 50 deg/day. The dominant cause of these non-detections, as already mentioned, is trailing loss: the reduction in sensitivity (i.e., brightening of the effective limiting magnitude) that all surveys experience for long-trailed NEOs relative to stationary or slow-moving objects. At 100 deg/day, an asteroid leaves a 125-arcsecond trail on a survey image taken with a 30-second exposure. At apparent magnitude 20, its trail intensity would be 24.5 mag per 2-arcsecond portion of trail. No currently operating survey is intended to detect such objects or capable of doing so --- although major advances in digital/synthetic tracking \citep[e.g.][]{Zhai2014,Heinze2015} could change this. 

\medskip

\begin{figure}
\includegraphics[width=3.5in]{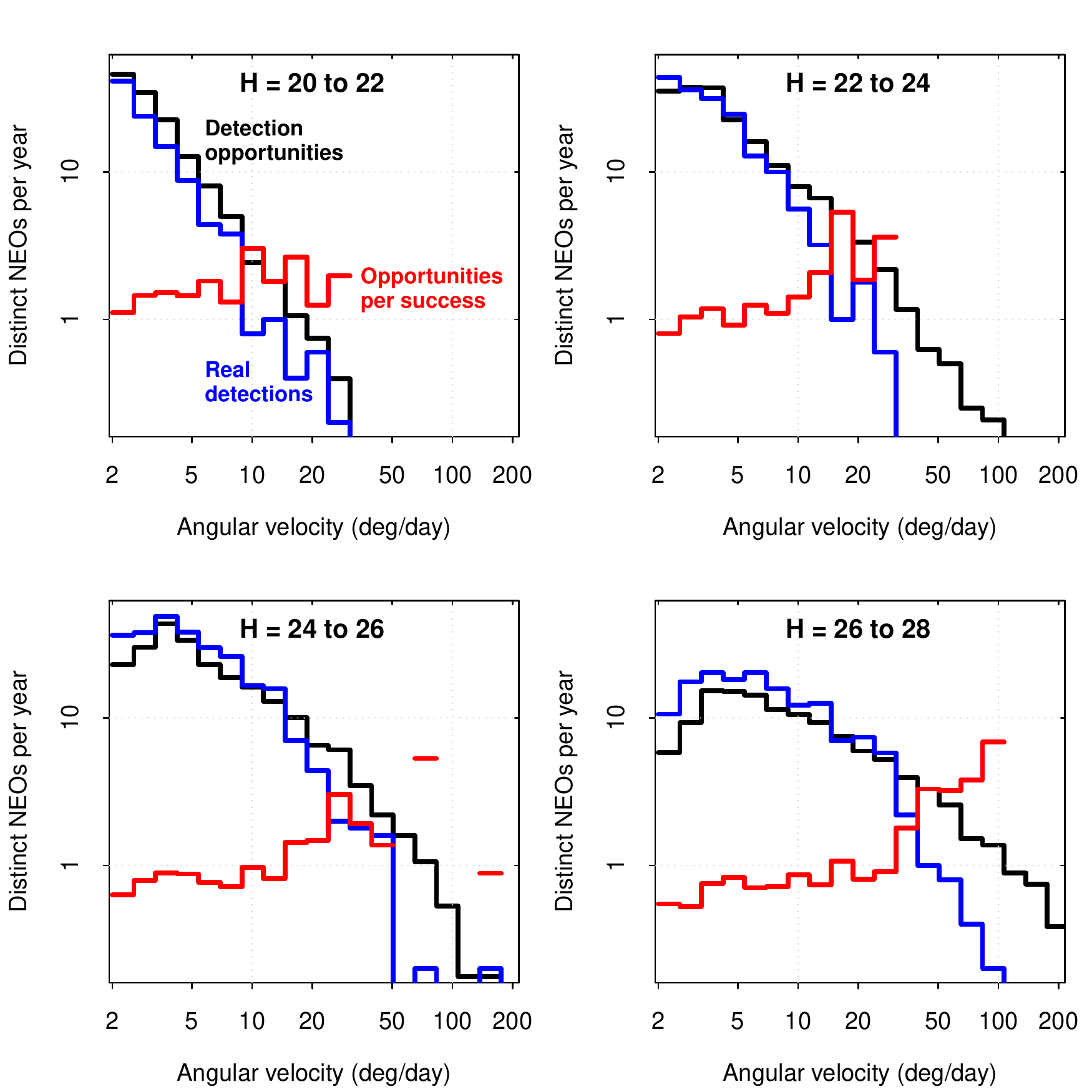}
\caption{Comparison of observed angular velocity distributions for NEOs vs. those predicted based on the orbital distribution from the Granvik model, combined with our results on the $H$ mag distribution, as parametrized by the power law fits from Figure \ref{fig:diffhist}. In contrast to Figure \ref{fig:angvelpred01} we have attempted to remove the effect of trailing losses by imposing a threshold trail intensity. Nevertheless, there is a clear signature of more detection opportunities being missed at higher angular velocities --- especially for the smallest asteroids. \label{fig:angvelpred02}}
\end{figure}

Huge trailing losses make it completely unreasonable to expect the faster-moving asteroids in the left panel of Figure \ref{fig:angvelpred01} to be detected, but are trailing losses solely responsible for the velocity bias we have observed? To explore this question, we attempt to construct new simulated histograms like those in Figure \ref{fig:angvelpred01}, with the effects of trailing loss accounted for. We re-sample the simulated ephemerides at an interval of 48 hours rather than 1 hour, to reduce the advantage the simulated asteroids enjoy of being `observed' at exactly the best moment, and then impose a threshold on trail intensity of magnitude 21.0 per 2 arcseconds. Thus, the `faintest' simulated asteroid permitted in the new analysis has an apparent magnitude of 20.0 and a trail intensity of 21.0. These values are slightly beyond ATLAS capability, but within the capacity of the other major surveys. Hence, our thresholds eliminate most simulated asteroids whose trailing losses are severe enough to prevent their detection in real life. Figure \ref{fig:angvelpred02} compares the resulting angular velocity histograms with the distributions actually observed, where the same thresholds of total brightness and trail intensity have been applied to the real tracklets.

\medskip

The comparisons in Figure \ref{fig:angvelpred02} are imprecise (because of the lack of color corrections, the arbitrary 48 hr sampling used for the simulation, and other effects). Nevertheless, there is a clear pattern of near one-to-one correspondence between predicted and actual NEO detections at slow angular velocities, grading into an increasing number of missed detection opportunities at higher angular velocities --- and the severity of this effect increases for the smaller asteroids.

\medskip

While definitive conclusions cannot be drawn from the imprecise comparison, Figure \ref{fig:angvelpred02} suggests that trailing losses alone do not account for the bias against detecting NEOs with high encounter velocities. Other effects likely contribute. These may include the shortness of the temporal window for observing a high-velocity asteroid; the difficulty of linking successive measurements due to the large departure from Great Circle motion; and the difficulty of following up such detections due to the rapidly increasing ephemeris uncertainty. In contrast to pure trailing losses, these effects can apply to NEOs with high encounter velocities even if their angular velocities are low because they are incoming for extremely close encounters or impacts.

\subsection{Risk Retirement}

Besides the potential immediate danger of failing to detect a high-velocity impactor, the bias against NEOs with fast encounter velocities is problematic in the context of estimating and retiring long-term impact risk. NEOs in orbits that pass Earth closely at high encounter velocity are disproportionately dangerous because their impacts would be more energetic. At the same time, they are under-represented in our discovery statistics, even for $H<23$. This means the fraction of un-retired risk for asteroids in a given size range is likely to be higher than the fraction of such objects remaining to be discovered, since the yet-to-be discovered objects are likely to have faster typical encounter velocities. 

\medskip

On the other hand, the impact cross section is larger for asteroids with low encounter velocities because of gravitational focusing: that is, Earth's gravity can pull an asteroid in to impact from a larger initial impact parameter if the asteroid is moving slowly. We will use $v_{\infty}$ to refer to the Earth-relative velocity of the asteroid while its distance is still much greater than $R_E$ (the radius of the Earth), and $b_{\infty}$ will be the asteroid's impact parameter with Earth if its trajectory were extrapolated without accounting for Earth's gravity. Note that, compared to $R_E$, the simulated and real asteroid encounters discussed herein are so distant that the encounter velocity $v_E$ and encounter distance $d_E$ are well approximated by $v_{\infty}$ and $b_{\infty}$: i.e., the Earth's gravity has relatively little effect on the asteroid trajectories. On the other hand, if the asteroid actually is going to hit the Earth, the maximum value of $b_{\infty}$ consistent with impact is:

\begin{equation} 
b_{\infty} = R_E \sqrt{1 + \frac{v_{esc}^2}{v_{\infty}^2}} \label{eq:gravfoc01}
\end{equation}

Where $R_E$ is the radius of the Earth and $v_{esc}$ is the escape velocity from Earth's surface (about 11.2 km/sec). The ratio of Earth's impact cross section to its geometrical cross section is then:

\begin{equation} 
\frac{b_{\infty}^2}{R_E^2} = 1 + \frac{v_{esc}^2}{v_{\infty}^2} \label{eq:gravfoc02}
\end{equation}

\begin{figure}
\includegraphics[width=3.5in]{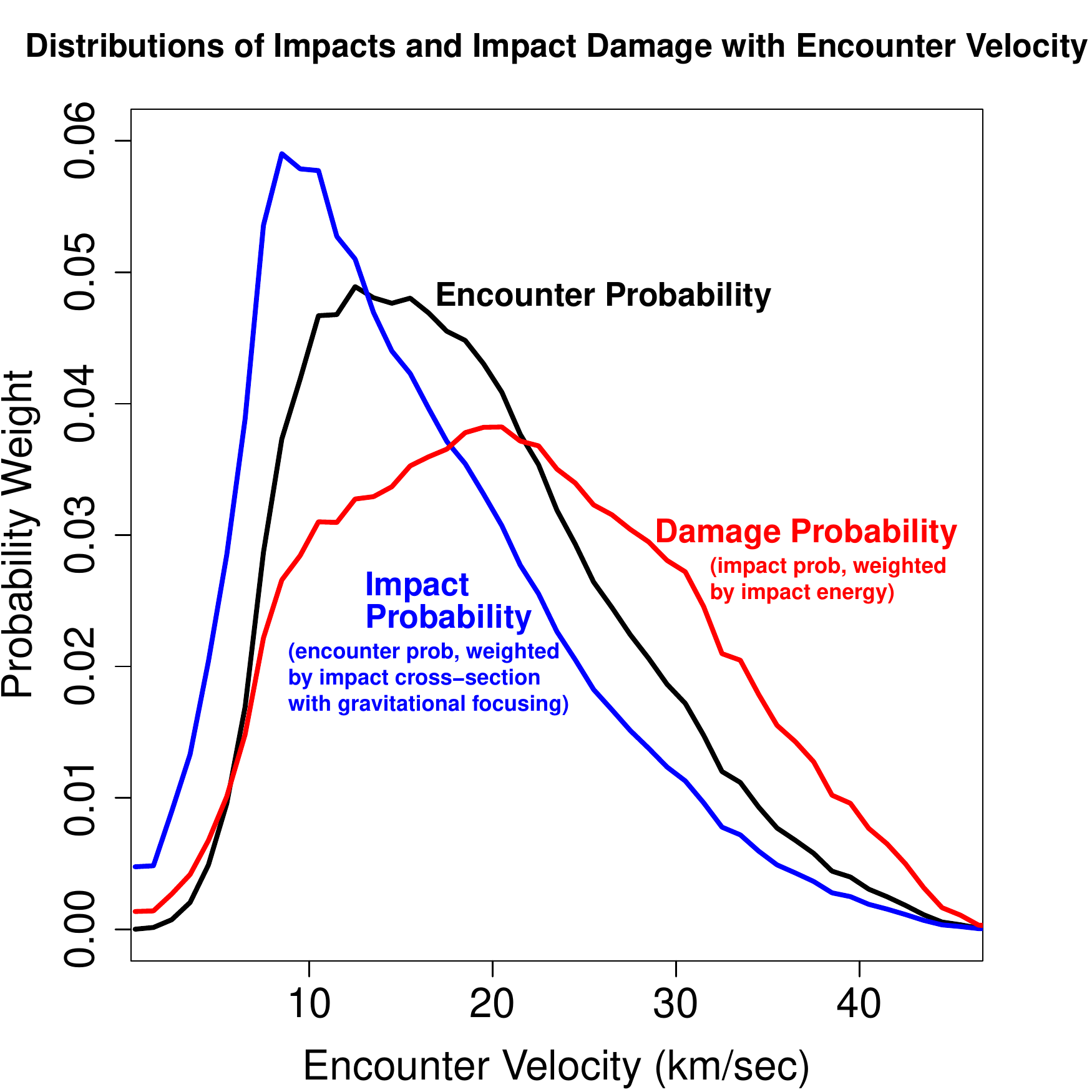}
\caption{Probability distributions over encounter velocity (which should be read as $v_{\infty}$ in the case of impactors; see text) for Earth encounters, Earth impacts, and damage due to impact. Relative to the encounter distribution, the impactors are weighted toward lower velocities because low-velocity asteroids have a larger impact cross section due to gravitational focusing (Equation \ref{eq:gravfoc02}). However, the damage is still strongly weighted toward higher-velocity asteroids because their impacts are so much more energetic.
\label{fig:damage}}
\end{figure}

This ratio of cross sections gets considerably larger than 1.0 for slow-velocity asteroids, and hence gravitational focusing is important in calculating impact cross sections and impact risk. Using the approximation that $v_E = v_{\infty}$ for (relatively) distant encounters, we can weight our previously-calculated histogram of encounter velocities by the velocity-dependent impact cross section from Equation \ref{eq:gravfoc02} to account for the fact that gravitational focusing makes slow-moving asteroids more likely to impact Earth. This weighted histogram is effectively a probability distribution showing how the risk of impact is distributed in terms of $v_{\infty}$. To account for the fact that higher-velocity asteroids, though less likely to impact, are more dangerous when they do, we can weight the probability distribution of impact risk by the impact energy, which is proportional to $v_{\infty}^2 + v_{esc}^2$. This produces a model of how the expected {\em damage} from asteroid impacts is distributed with $v_{\infty}$, assuming that damage is proportional to impact energy. As illustrated by Figure \ref{fig:damage}, the strong positive dependence of impact energy on encounter velocity overwhelms the opposing effect of gravitational focusing: the expected damage from Earth impacts is weighted toward asteroids with higher encounter velocities. Hence, this analysis substantiates our earlier claim that a disproportionate fraction of planetary risk resides in the higher-velocity asteroids that are less likely to have yet been discovered.

\section{Conclusion} \label{sec:conc}

We have attempted to assess the total NEO population of the Solar System using a combination of observations and simulation that is in the conceptual tradition of previous work \citep[e.g.][]{Harris2015,Lilly2017,Stokes2017,Tricarico2017,Trilling2017}, but is the first to use data from the Asteroid Terrestrial-impact Last Alert System \citep[ATLAS; ][]{Tonry2018}, and the first to adopt a comprehensive `Solar System to pixels' simulation strategy for debiasing the observed asteroid counts. 

\medskip

Our population estimates fall in the range of previous results, though somewhat below the median at most $H$ magnitudes. Most errors to which our calculation is susceptible would tend to {\em underestimate} the true number of NEOs, so the actual population may be closer to some of the higher estimates.

\medskip

Consistent with the Granvik model and several observational results, we find evidence for a significant change in slope of the NEO size distribution between $H=22$ and $H=23$ (e.g., at a size of about 100 meters). This probably corresponds to the transition between larger objects whose internal cohesion is mainly due to self-gravity and smaller asteroids held together by material strength \citep[e.g.][]{Harris2015}. For $H \geq 23$, we can fit the differential size distribution (Equation \ref{eq:pwr02}) with a power law of slope $b_d \sim 3.8$. Though this slope value is very tentative due to uncertainties from `lost' NEOs, it is close to $b_d = 4.0$, the boundary beyond which cumulative mass diverges as size goes to zero. This may indicate that very small NEOs make a relatively large contribution to the population's total mass budget. 

\medskip

We find a strong bias against the detection of NEOs that encounter Earth with large relative velocities. This bias is especially strong for asteroids smaller than $H\sim23$ (e.g. size about 100 meters), and is common to the ATLAS simulation; real ATLAS data; and global NEO discovery statistics. The bias is likely caused by a combination of factors: 

\begin{enumerate}

\item NEOs that encounter Earth with fast relative velocities tend to have higher angular velocities on the sky: hence, they appear on survey images as long, faint trails (i.e., they suffer from severe trailing losses).

\item High encounter velocity translates into a more rapid increase in ephemeris uncertainty following an initial detection, which makes successful followup observations less likely.

\item NEOs with high encounter velocities tend to have greater angular accelerations in their on-sky motion, which makes more difficult both the linking of successive detections and the targeting of followup observations.

\item High encounter velocity means an asteroid spends a relatively brief time close to Earth: there is only a short window of opportunity when its discovery is possible.

\end{enumerate}

\medskip

The first factor listed above is a direct result of fast angular velocity, and would not apply to NEOs that have extremely close approaches (e.g. 0.001 AU) or impacts with Earth. This is because such objects approach and recede from Earth along nearly radial trajectories, and hence have low angular velocities except for a very brief period surrounding the moment of closest approach. If angular velocity is the dominant cause of the bias against detection of NEOs with high encounter velocities, the bias should disappear for the very closest encounters. On the other hand, if the other factors listed above are important contributors, the bias against high encounter velocities may persist even for impactors, making it more of a concern for planetary defense. Our analysis of the angular velocity distributions in Section \ref{sec:angveldist} seems to suggest trailing losses do not explain the entirety of the bias: hence, the other, more concerning effects may be important.

\medskip

ATLAS is actively working on software to detect and link longer-trailed NEOs in spite of Great Circle residuals. Such improvements are relevant to items one and three in the list above. Item four can be addressed only by building more telescopes in more geographically diverse locations. This is important not only to attain round-the-clock coverage of both celestial hemispheres; but also for weather diversity --- that is, to ensure that localized periods of bad weather cannot greatly reduce global ability to detect incoming impactors. In aid of this mission, ATLAS is currently constructing new survey units in South Africa and Chile. 

\medskip

Regardless of whether the bias against detecting NEOs with high encounter velocities persists for impactors, it can be reduced by prompt submission and rapid, aggressive followup for NEO candidates --- improvements which address both the second and the third items in the list above. Most candidates {\em are} successfully followed up at present, but there is reason to believe \citep[e.g.][and our analysis of lost ATLAS objects and worldwide close-approach statistics]{Veres2018} that some of the most interesting objects with fast velocities and close encounter distances are still being lost. In the future, increased discoveries and reduced losses may result from the coming online of the Vera C. Rubin Observatory (formerly LSST), but this will depend on well-designed data mining and followup endeavors.

\medskip

The prompt and aggressive followup needed to reduce NEO losses (with or without the Rubin Observatory) can be fostered and leveraged through closer collaboration between the various surveys, between surveys and the global followup community, and between observers and orbital analysts. Efforts along these lines are already underway, and we hope our analysis will encourage those involved as to their value and importance.

\section{Acknowledgments}

Support for the ATLAS survey was provided by NASA grant NN12AR55G under the guidance of Lindley Johnson and Kelly Fast. This research has made use of data and/or services provided by the International Astronomical Union's Minor Planet Center. This publication makes use of data products from the Pan-STARRS1 Surveys and the PS1 public science archive, which have been made possible through contributions by the Institute for Astronomy, the University of Hawaii, the Pan-STARRS Project Office, the Max-Planck Society and its participating institutes, the Max Planck Institute for Astronomy, Heidelberg and the Max Planck Institute for Extraterrestrial Physics, Garching, The Johns Hopkins University, Durham University, the University of Edinburgh, the Queen's University Belfast, the Harvard-Smithsonian Center for Astrophysics, the Las Cumbres Observatory Global Telescope Network Incorporated, the National Central University of Taiwan, the Space Telescope Science Institute, the National Aeronautics and Space Administration under Grant No. NNX08AR22G issued through the Planetary Science Division of the NASA Science Mission Directorate, the National Science Foundation Grant No. AST-1238877, the University of Maryland, Eotvos Lorand University (ELTE), the Los Alamos National Laboratory, and the Gordon and Betty Moore Foundation.


\begin{thebibliography}{}
\bibitem[Alard \& Lupton (1998)]{Alard1998} Alard, C. \& Lupton, R. H. 1998, \apj, 503, 325
\bibitem[Alard (2000)]{Alard2000} Alard, C. 2000, \aaps, 144, 363
\bibitem[Alcock et al.(1999)]{Alcock1999} Alcock, C., Allsman, R. A., Alves. D., et al. 1999, \apj, 521, 602
\bibitem[Akerlof et al.(2000)]{ROTSE} Akerlof, C., Alexandroff, F., Allende Prieto, C., et al. 2000, \aj, 119, 1901
\bibitem[Becker (2015)]{hotpants} Becker, A. 2015, HOTPANTS: High Order Transform of PSF ANd Template Subtraction, Astrophysics Source Code Library, asc:1504.004
\bibitem[Bottke et al.(2002)]{Bottke2002} Bottke, W. F., Morbidelli, A., Jedicke, R., Petit, J.-M. , Levision, H. F., Michel, P. \& Metcalfe T. S. 2002, Icarus, 156, 399
\bibitem[Bottke et al.(2005)]{Bottke2005} Bottke, W. F., Durda, D. D., Nesvorn\'{y}, D., Jedike, R., Morbidelli, A., Vokrouhlick\'{y}, D., \& Levison, H. F. 2005, Icarus, 179, 63
\bibitem[Bowell et al.(1989)]{Bowell1989} Bowell, E., Hapke, B., Domingue, D., et al. 1989, Asteroids II (Tucson, AZ: Univ. Arizona Press), 524
\bibitem[de El\'{i}a \& Brunini (2007)]{deElia2007} de El\'{i}a, G. C \& Brunini, A. 2007, \aap, 466, 1159
\bibitem[Denneau et al.(2013)]{Denneau2013} Denneau, L., Jedicke, R., \& Grav, T. et al. 2013, \pasp, 125, 357
\bibitem[Dunham et al.(2013)]{Dunham2013} Dunham, D. W., Reitsema, H. J., Lu, E., Arentz, R., Linfield, R., Chapman, C., Farquhar, R., Ledkov, A. A., Eismont, N. A., \& Chumachenko, E. 2013, SoSyR, 47, 315
\bibitem[Erasmus et al.(2017)]{Erasmus2017} Erasmus, M., Mommert, M., Trilling, D.E., Sikafoose, A. A., van Gend, C., \& Hora, J. L. 2017, \aj, 154, 162
\bibitem[Erasmus et al.(2018)]{Erasmus2018} Erasmus, M., McNeill, A., Mommert, M., Trilling, D.E., Sikafoose, A. A. \& van Gend, C. 2018, \apjs, 237, 19
\bibitem[Farinella et al.(1998)]{Farinella1998} Farinella, P., Vokrouhlick\'{y}, D. \& Hartmann, W. K. 1998, Icarus, 132, 378
\bibitem[Gladman et al.(2009)]{Gladman2009} Gladman, B. J., Davis, D. R., Neese, C., Jedicke, R., Williams, G., Kavelaars, J. J., Petit, J-M., Scholl, H., Holman, M., Warrington, B., Esquerdo, G., \& Tricarico, P. 2009, Icarus, 202, 104
\bibitem[Granvik et al.(2018)]{Granvik2018} Granvik, M., Morbidelli, A., Jedicke, R., Bolin, B., Bottke, W. F., Beshore, E., Vokrouhlick\'{y}, D., Nesvorn\'{y}, D., Michel, P. 2018, Icarus, 312, 181
\bibitem[Harris \& D'Abramo (2015)]{Harris2015} Harris, A. W. \& D'Abramo, G. 2015 2015, \icarus, 257, 302
\bibitem[Heinze et al.(2015)]{Heinze2015} Heinze, A. N., Metchev, S., \& Trollo, J. 2015, \aj, 150, 125
\bibitem[Larson et al.(2003)]{CSS} Larson, S., Beshore, E., Hill, R., et al. 2003, DPS, 35, 3604
\bibitem[Lucy (1974)]{RLdeconL} Lucy, L. B. 1974, \aj, 79, 745
\bibitem[Marshall et al.(2017)]{LSST} Marshall, P. et al. 2017, arXiv:1708.04058
\bibitem[Mommert et al.(2016)]{Mommert2016} Mommert, M., Trilling, D. E., Borth, D., Jedicke, R., Butler, N., Reyes-Ruiz, M., Pichardo, B., Petersen, E., Axelrod, T., \& Moskovitz, N. 2016, \aj, 151, 98
\bibitem[Morbidelli et al.(2020)]{Morbidelli2020} Morbidelli, A., Delbo, M., Granvik, M., Bottke, W. F., Jedicke, R., Bolin, B., Michel, P., \& Vokkrouhlicky, D. 2020, \icarus, 340, 113631
\bibitem[Nesvorn\'{y} \& Bottke (2004)]{Nesvorny2004} Nesvorn\'{y}, D. \& Bottke, W. F. 2004, Icarus, 170, 324
\bibitem[Pravec et al.(2012)]{Pravec2012} Pravec, P., Harris, A., Ku\v{s}nir\'{a}k, P., Gal\'{a}d, A., \& Hornoch, K. 2012, \icarus, 221, 365
\bibitem[Press et al.(1992)]{nrc} Press, W. H., Teukolsky, S.A., Vetterling, W. T., \& Flannery, B. P. 1992, Numerical Recipes in C (Second Edition; New York, NY: Cambridge University Press)
\bibitem[Richardson (1972)]{RLdeconR} Richardson, W. H., 1972, J. Opt. Soc. Am. 62, 55
\bibitem[Stokes et al.(2000)]{LINEAR} Stokes, G. H., Evans, J. B., Viggh, H. E. M., Shelly, F. C. \& Pearce, E. C. 2000, Icarus, 148, 21
\bibitem[Schunov\'{a}-Lilly et al.(2017)]{Lilly2017} Schunov\'{a}-Lilly, E., Jedicke, R., Vere\v{s}, Denneau, L., \& Wainscoat, R. 2017, \icarus, 284, 114
\bibitem[Smith et al.(2020)]{Smith2020} Smith, K. W., Smartt, S. J., Young, D. R. 2020, \pasp, 132, 085002
\bibitem[Stokes et al.(2017)]{Stokes2017} Stokes, G., Barbee, B. W., Bottke, W. F., et al. 2017, Update to Determine the Feasibility of Enhancing the Search and Characterization of NEOs, Report of the Near-Earth Object Science Definition Team(Washington, D.C:Science Mission Directorate, Planetary Science Division, NASA)
\bibitem[Tonry et al.(2018a)]{Tonry2018} Tonry, J. L., Denneau, L., Heinze, A. N., Stalder, B., Smith, K. W., Smartt, S. J., Stubbs, C. W., Weiland, H. J., \& Rest, A. 2018, \pasp, 130, 4505
\bibitem[Tonry et al.(2018b)]{refcat} Tonry, J. L., Denneau, L., Flewelling, H., Heinze, A. N., Onken, C. A., Smartt, S. J., Stalder, B., Weiland, H. J. \& Wolf, C. 2018, \apj, 867, 105
\bibitem[Tricarico (2017)]{Tricarico2017} Tricarico, P. 2017, \icarus, 284, 416
\bibitem[Trilling et al.(2017)]{Trilling2017} Trilling, D. E., Valdes, F., Allen, L., James, D., Fuentes, C., Herrera, D., Axelrod, T., \& Rajagopal, J. 2017, \aj, 154, 170
\bibitem[Wiegert et al.(2007)]{Wiegert2007} Wiegert, P., Balam, D., Moss, A., Veillet, C., Connors, M., \& Shelton, I. 2007, \aj, 133, 1609
\bibitem[Vere\v{s} et al.(2015)]{Veres2015} Vere\v{s}, P., Jedicke, R., Fitzsimmons, A., Denneau, L., Granvik, M., Bolin, B., Chastel, S., Wainscoat, R. J., Bergett, W. S., Chambers, K. C., Flewelling, H., Kaiser, N., Magnier, E. A., Morgan, J. S., Price, P. A., Tonry, J. L., \& Waters, C. 2015, \icarus, 261, 34
\bibitem[Vere\v{s} et al.(2018)]{Veres2018} Vere\v{s}, P., Payne, M. J., Holman, M. J., Farnocchia, D., Williams, G. V., Keys, S., \& Boardman, I. 2018, \aj, 156, 5
\bibitem[Ye et al.(2019)]{Ye2019} Ye, Q., Masci, F. J., Lin, H. W. et al. 2019, \pasp, 131, 078002
\bibitem[Yoshida \& Nakamura (2007)]{Yoshida2007} Yoshida, F. \& Nakamura, T. 2007, \planss, 55, 1113
\bibitem[Zhai et al.(2014)]{Zhai2014} Zhai, C., Shao, M., Nemati, B., Werne, T., Zhou, H., Turyshev, S. G., Sandhu, J., Hallinan, G., \& Harding, L. K. 2014, \apj, 792, 60
\end{thebibliography}
\end{document}